\documentclass[twocolumn,superscriptaddress,showpacs,amsmath,amssymb,aps,pre,floatfix,prstab,prstper,nolongbibliography]{revtex4-1}

\usepackage{color}    
\usepackage{graphicx}
\usepackage{braket} 
\usepackage{dcolumn}
\usepackage{bm}
\usepackage{subfigure}
\usepackage{amssymb}
\usepackage{multirow}
\usepackage{natbib}
    \usepackage{amsmath}
\graphicspath{{plots/}}
\usepackage[english]{babel}

\usepackage[colorlinks,linkcolor=blue,anchorcolor=blue,citecolor=blue,urlcolor=blue,CJKbookmarks=True]{hyperref}
\begin{document}


\title{Finite temperature density functional theory investigation to the nonequilibrium transient warm dense state created by 
laser excitation}


	\author{Hengyu~Zhang}
	\affiliation{Department of Physics, National University of Defense Technology, Changsha, Hunan 410073, China}
	\author{Shen~Zhang}
	\email{shenzhang@nudt.edu.cn}
	\affiliation{Department of Physics, National University of Defense Technology, Changsha, Hunan 410073, China}
	\affiliation{Institut f\"ur Theoretische Physik und Astrophysik, Christian-Albrechts-Universit\"at zu Kiel,
		Leibnizstra{\ss}e 15, 24098 Kiel, Germany}
	\author{Dongdong Kang}
	\affiliation{Department of Physics, National University of Defense Technology, Changsha, Hunan 410073, China}
	\author{Jiayu~Dai}
	\email{jydai@nudt.edu.cn}
	\affiliation{Department of Physics, National University of Defense Technology, Changsha, Hunan 410073, China}
	\author{M.~Bonitz}
	\affiliation{Institut f\"ur Theoretische Physik und Astrophysik, Christian-Albrechts-Universit\"at zu Kiel,
		Leibnizstra{\ss}e 15, 24098 Kiel, Germany}

\begin{abstract}
We present a finite-temperature density functional theory investigation of the nonequilibrium transient electronic structure of warm dense Li, Al, Cu, and Au created by laser excitation.
Photons excite electrons either from the inner shell orbitals or from the valence bands according to the photon energy, 
and give rise to isochoric heating of the sample. Localized states related to the 3d orbital are observed for Cu when the hole lies in the inner shell 3s orbital. The electrical conductivity for these materials at nonequilibrium states is calculated using the Kubo-Greenwood formula. The change of the electrical conductivity, compared to the equilibrium state, is different for the case of holes in inner shell orbitals or the valence band. This is attributed to the competition of two factors: the shift of the orbital energies due to reduced screening of core electrons, and the increase of chemical potential due to the excitation of electrons. The finite temperature effect of both the electrons and the ions on the electrical conductivity is discussed in detail. This work is helpful to better understand the physics of laser excitation experiments of warm dense matter.
\end{abstract}








\maketitle

\section{Introduction}\label{s:intro}

Warm dense matter (WDM), referring to the intermediate state between condensed matter and ideal plasmas, has attracted numerous studies in recent years. It is not only because WDM bridges the gap between atomic physics, condensed matter physics and plasmas physics, but also because these extreme WDM conditions are found in a wide range of applications~\cite{graziani-book,bonitz_pop_20}. WDM can be characterized by the ion coupling parameter, $\Gamma = e^2/(\bar r k_BT)\sim $ 1, the electron degeneracy parameter $\Theta=k_BT/E_F \sim$ 1, and the quantum coupling parameter $r_s=\bar r/a_B \sim 1$, where $\Gamma$ is the ratio of the Coulomb energy to the thermal energy, $\Theta$ is the ratio of the thermal energy to the Fermi energy, and $r_s$ is the ratio of the mean interparticle distance to the Bohr radius, respectively. The WDM regime typically covers a wide range of temperature from several thousand Kelvin to tens of millions of Kelvin, and a density range from near solid density to a thousand times of solid density. The properties of WDM, such as equation of state (EOS), thermal and electrical conductivity, etc., are of great importance to model the structure and understand the formation of celestial bodies~\cite{PhysRevResearch.2.023260,doi:10.1002/ctpp.2150350203,extFPMD,Lu_2019}, such as the earth, brown dwarfs and giant planets. In addition, the deuterium-tritium fuel for inertial confinement fusion (ICF) passes through the WDM regime before ignition~\cite{PhysRevE.89.043105,doi:10.1063/5.0008231}.  
 
One of the widely used techniques to produce WDM in the laboratory is isochoric heating of a solid sample, e.g.~\cite{falk_2018}, with the choice of available sources such as optical lasers~\cite{PhysRevLett.82.4843,PhysRevLett.110.135001,doi:10.1063/1.2816439}, ion beams~\cite{PhysRevLett.91.125004,Roth_2009,MANCIC201021}, or X-ray free electron lasers (XFEL)~\cite{Vinko2012,xfel_carbon,Kraus_2018}. With the development of intense ultrafast optical lasers, the intensity can be as high as 10$^{22}$ W/cm$^2$ and the pulse duration can be less than 50 fs~\cite{chen_ng,Hengyu,RevModPhys.84.1177,Kiriyama}. During the pulse, the laser energy is transferred to the electrons via photoexcitation of valence electrons to the conduction band with negligible movement of the much heavier ions. In order to ensure spatial homogeneity, the thickness of the target sample is usually very small, typically on the order of 10 nm. The intensity of the laser beam is usually on the order of 10$^{13}$ W/cm$^2$, for such isochoric heating experiments. Compared to high intensity ultrafast optical lasers, the XFEL has a much higher photon energy, which delivers the XFEL energy to the sample via photoionization of core electrons. XFEL radiation can also penetrate deeper into the sample (on the order of 10 $\mu$m), and the intensity of the beam can be as high as 10$^{20}$ W/cm$^2$~\cite{mimura2014generation}, allowing even severe depletion of certain inner shells~\cite{Nagler2009,Vinko2012}. The holes that are created by photoexcitation or photoionization will be refilled via Auger decay or other collisional processes. However, the lifetime of such a vacancy is typically comparable to the  pulse duration (for example, the lifetime of a hole in the L shell is estimated to be in the range of 10 fs, for rare gas atoms ~\cite{schuette_prl_12}, 40 fs for Al~\cite{PhysRevB.39.3489}, and 34.5 fs for Cu~\cite{PhysRevLett.83.832}.), which is short compared with any electron-phonon coupling time (on a picosecond scale)~\cite{PhysRevLett.122.015001}. As a result, the isochoric heating technique can create a variety of nonequilibrium transient states of matter in the laboratory including nonequilibrium electronic density of states and non-thermal melting, e.g. \cite{sokolowski_99}.

On the other hand, theoretical investigation of such a nonequilibrium transient state faces great challenges. Atomic kinetic simulations are widely used to interpret the experimental results~\cite{SCFLY,medvedev_10}, but the quality of their parameters under such exotic conditions is not well examined. Nonequilibrium Green functions are a powerful tool to study the ultrafast dynamics of scattering processes in atoms and condensed matter, e.g. \cite{balzer_pra_10_2,hermanns_prb14,schluenzen_jpcm_19}, but the high computational cost allows to study only simple model systems.
Time-dependent DFT (TD-DFT) is a feasible theoretical tool for the investigation to WDM~\cite{Baczewski, PhysRevLett.120.205002} including the nonequilibrium dynamics, because it explicitly solves the time-dependent Kohn-Sham equations to obtain the temporal evolution of the wave function~\cite{TDDFT}. The nonequilibrium dynamics of the electronic structure obtained by TD-DFT breaks down into two relaxation channels, i.e. the density of states reorganization and the redistribution of the electron number~\cite{silaeva2018}. However, TD-DFT calculation is also very expensive and the knowledge of its exchange-correlation functional is limited. In addition, the dynamics such as Auger decay and electron temperature cannot be treated directly in TD-DFT. 

Here we concentrate on a simpler and much faster approach -- finite-temperature density functional theory (FTDFT). It does not rely on any fitting parameters from the experiment and has been widely used to study the properties of WDM~\cite{PhysRevB.93.115114,PhysRevLett.104.245001,Dai_2017}. However, FTDFT is constructed for the calculation of equilibrium states, and the presence of a hole is usually accounted for in FTDFT  by a specially designed pseudopotential with the configuration of an ionized atom~\cite{RN100}. The hole is, therefore, fixed in the frozen core of a particular atom serving as an impurity in the system~\cite{taillefumier2002,RN98}. 

In this work, we have modified the original Kohn-Sham-Mermin scheme to allow for a nonequilibrium electron distribution function within FTDFT to study transient exotic states that were described above. With the manually introduced redistribution of the electron occupations, this modified FTDFT treatment of nonequilibrium states has a much better efficiency compared to the more advanced TD-DFT counterpart. The electronic structures of the metallic systems of lithium, aluminum and copper with a variable number of holes in their inner shell orbitals are investigated theoretically by the modified code, corresponding to the nonequilibrium transient state immediately after 
photoionization
by an XFEL pulse. For comparison, we also present the electronic structure of copper and gold with variable numbers of holes in the valence band, which can be realized via photoexcitation by means of a high intensity ultrafast laser. The electrical conductivity -- a widely used diagnostic technique for WDM~\cite{Celliers677} -- is also calculated to demonstrate the effect of such holes on transport properties. Clearly such a treatment neglects relaxation processes and gives, at best, only intermediate states of the system after excitation. Nevertheless, we expect that this approach allows one to study important trends of nonequilibrium behavior which has been confirmed in experiments \cite{Nagler2009}. Support for such a separation of relaxation phases is also based on similar approaches in other fields including the three-step model of photoionization, e.g. \cite{schattke-book} and the modeling of so-called hollow atoms where core electrons are removed by the impact of a highly charged ion, e.g. \cite{aumayr_99}.

The paper is organized as follows: in Sec. \ref{s:theory}, the theoretical method together with the numerical details used in this work are described. In Sec. \ref{s:results}, results and discussions are presented including the electronic structure and electrical conductivity of metallic systems with variable numbers of holes in both, inner shell and valence band, and for electron temperatures of 300K and 1eV, respectively. Our conclusions are summarized in Sec.~\ref{s:conclusion}.

\section{METHODOLOGY AND NUMERICAL DETAILS}\label{s:theory}
 In this section we describe the methodology and numerical details of this work. FTDFT calculations are performed with the population of electrons manually controlled in the self-consistent field (SCF) cycle to create holes in different bands. Then the Kubo-Greenwood formula is used to calculate the electrical conductivity with the wave functions obtained by the FTDFT calculation.  
 
\subsection{Finite temperature density functional theory and molecular dynamics} \label{sa:dft}
Since the energy transfer between the electrons and the ions is much slower than the evolution of the electron subsystem, a two-temperature system~\cite{anisimov1974electron,RN99,PhysRevB.77.075133} can be used, where the electrons are described by FTDFT with a fixed ion configuration. FTDFT calculations are performed using the plane-wave-pseudopotential open-source package \textsc{Quantum ESPRESSO} (QE)~\cite{giannozzi2009,Giannozzi_2017} with minor modifications to realize the manual control of the  occupation numbers. In the Kohn-Sham-Mermin scheme~\cite{KohnSham, Mermin}, the electrons are considered to be in thermal equilibrium and their occupation numbers follow the Fermi-Dirac distribution. In order to investigate the nonequilibrium state such as the one of the WDM system created by isochoric heating, we force the occupation numbers of the relevant orbitals to be zero in every SCF step, which manually creates holes in the inner shell orbitals or the valence band. The total number of electrons in the calculation system remains unchanged by adding the same number of electrons as that of holes to the chemical potential. Similar to the original SCF process, the new chemical potential is computed self-consistently, which, in return, determines the occupation numbers of all Kohn-Sham orbitals, except for those that were forced to zero. By this way, we can go beyond the frozen core assumption and create holes in the valence band directly in our DFT calculations, contrary to the  common practice of creating holes during the generation of pseudopotentials and treating the corresponding atom as an impurity in the DFT calculation. 

We consider the metallic systems of lithium, aluminum, copper and gold as illustrating examples for our nonequilibrium FTDFT calculations. Cubic boxes with periodic boundary conditions are used for the 4 elements mentioned above, and the system sizes are set to 128 atoms for body-centered cubic (BCC) Li, and 108 atoms for the remaining face-centered cubic (FCC) materials to guarantee the convergence with respect to finite size effect. Except for the molecular dynamics calculations, the atoms are fixed in their perfect lattice configuration with varying electronic temperature. We use the projector augmented wave (PAW) formalism~\cite{PAW} for the FTDFT calculation with the local density approximation (LDA) \cite{PerdewZunger} to the exchange-correlation interaction throughout. Finite temperature effects in the exchange-correlation functional are examined by the comparison with the newly constructed GDSMFB functional~\cite{groth_prl17}. The minimum threshold of the occupation numbers is 10$^{-6}$, in all our calculations, to ensure the convergence for the number of included bands. 

For Li, both K-shell and L-shell electrons (i.e., 1s$^2$2s$^1$) are treated as valence electrons, and the plane wave cutoff energy is set to be 100 Ry. For Al, we use a pseudopotential with both L-shell and M-shell electrons (i.e., 2s$^2$2p$^6$3s$^2$3p$^1$) included as valence electrons with a plane wave cutoff energy of 100 Ry. For Cu, two different pseudopotentials are applied for comparison. One includes 11 electrons as the valence electrons (i.e., 3d$^{10}$4s$^1$), and the other includes 19 electrons (i.e., 3s$^2$3p$^6$3d$^{10}$4s$^1$). We use 80 Ry for the plane wave cutoff energy, for the former pseudopotential, and 120 Ry, for the latter one. 11 valence electrons (i.e., 5d$^{10}$6s$^1$) are considered for the calculation of Au, and the cutoff energy for plane wave basis is 100 Ry. Unshifted Monkhorst-Pack K-point meshes are applied for the sampling of the Brillouin zone~\cite{MPgrid}, with $4\times4\times4$ size for Li, Al, and Cu, with 19 valence electrons, and a size of $8\times8\times8$, for 11-valence-electrons Cu and Au for the better description of the electronic structure near the chemical potential. The convergence with respect to both plane wave cutoff energy and K point grid size is carefully checked for all our calculations. 

To demonstrate the effect of the ionic structure on the electrical conductivity, we also perform FTDFT calculations combining with molecular dynamics (MD) simulations based on Born-Oppenheimer approximation~\cite{AIMD}. The FTDFT-MD calculation of Cu is carried out using a pseudopotential with 11 valence electrons, as an illustrating example. We apply a canonical (NVT) ensemble, where the ions, controlled by an Andersen thermostat, share the same temperature as the electrons. A time step of 1 fs is used, and the system is thermalized for more than 1 ps to reach equilibrium before the ionic trajectories of the last 2000 MD steps are kept for electronic structure calculation. 

\subsection{Kubo-Greenwood formula for electrical conductivity} \label{sa:kgec}
The electrical conductivity calculations are performed by using the KGEC code~\cite{calderin2017}. The wave functions with holes in the system obtained in the previous modified FTDFT calculation are plugged into the KGEC code with similar modification of manually controlled occupation numbers for the corresponding orbitals. 
Using the Kubo-Greenwood linear response formula, the real part of the electrical conductivity above the direct current (DC) limit can be calculated as 
\begin{equation}
\begin{split}
\label{sa:sigma1'}
    \sigma_{1}(\omega) = \frac{2{\pi}e^{2}\hbar^{3}}{m_{e}^{2}\Omega}\Sigma_{\textbf{k}}w_{\textbf{k}}\Sigma_{nn^{'}}
    \frac{\Delta{f}_{{n}^{'}\textbf{k},n\textbf{k}}}{\Delta\epsilon_{n\textbf{k},{n}^{'}\textbf{k}}}\langle\Psi_{n\textbf{k}}|\nabla|\Psi_{n^{'}\textbf{k}}\rangle \times\\
    \langle\Psi_{n^{'}\textbf{k}}|\nabla|\Psi_{n\textbf{k}}\rangle\delta(\Delta\epsilon_{n\textbf{k},{n}^{'}\textbf{k}}-\hbar\omega)\,,
\end{split}
\end{equation} 
which, near the DC limit, takes the following form
\begin{equation}
\begin{split}
\label{sa:sigma1''}
    \sigma_{1}(\omega) = \frac{2{\pi}e^{2}\hbar^{3}}{m_{e}^{2}\Omega\omega}\Sigma_{\textbf{k}}w_{\textbf{k}}\Sigma_{nn^{'}}\Delta{f}_{{n}^{'}\textbf{k},n\textbf{k}}\langle\Psi_{n\textbf{k}}|\nabla|\Psi_{n^{'}\textbf{k}}\rangle \\
    \langle\Psi_{n^{'}\textbf{k}}|\nabla|\Psi_{n\textbf{k}}\rangle\delta(\Delta\epsilon_{n\textbf{k},{n}^{'}\textbf{k}}-\hbar\omega).
\end{split}
\end{equation}
Here $\sigma_{1}$ is the real part of the electrical conductivity as a function of frequency $\omega$. The constants $e$, $\hbar$, $m_{e}$ and $\Omega$ represent the electron charge, Planck's constant, electron mass, and cell volume, respectively. The numbers $n$ and $n'$ denote the band index. Together with the Brillouin zone wave vector $\textbf{k}$ they become a pair index for Bloch states. The coefficient $w_k$ is the integration weight of the respective k-point, and $\Psi_{n\textbf{k}}$ is the wave function. $\Delta\epsilon_{n\textbf{k}, n'\textbf{k}} = \epsilon_{n\textbf{k}} - \epsilon_{n'\textbf{k}}$ and $\Delta{f}_{n'\textbf{k},n\textbf{k}} = {f}(\epsilon_{n'\textbf{k}}) - {f}(\epsilon_{n\textbf{k}})$ are the differences of Kohn-Sham eigenvalues and nonequilibrium occupation numbers obtained by the previous FTDFT calculation, respectively. $\delta$ is the Dirac delta function, which is broadened (we use a Lorentzian form) in all our calculations.

\section{Results and discussions}\label{s:results}
\subsection{Electronic structure calculations} \label{sa:dos}
\begin{figure}
    \centering
    \includegraphics[width=0.41\textwidth]{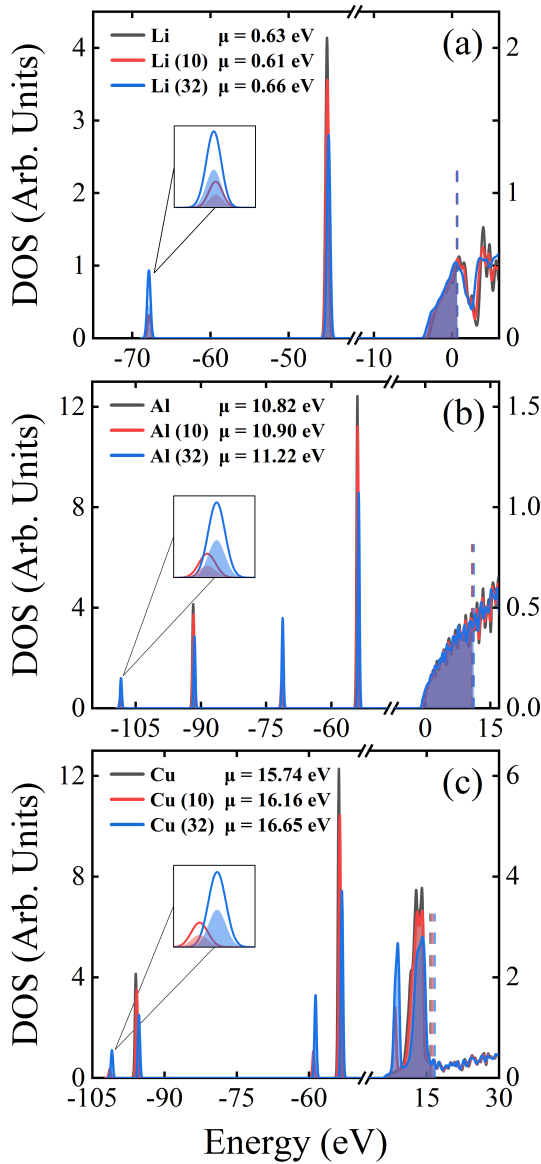}
    \caption{Density of states of (a) BCC Li, (b) FCC Al and (c) FCC Cu with a perfect lattice structure at ambient condition. The black solid lines show the calculated DOS with the occupied DOS shown as shades. The positions of their chemical potential are marked with the vertical dashed black lines and labelled in the legends. The red lines represent the same quantity as the black lines but are calculated with 10 electrons removed from the inner shell orbitals (1s for Li, 2s for Al, and 3s for Cu). For the blue lines, the number of such holes is increased to 32. For a better display, the DOS for the inner shell orbitals and for the valence band use different scales corresponding to the left and right axes.}
    \label{fig:Fig1}
\end{figure}

The electronic structure of Li, Al, Cu and Au with a nonequilibrium distribution of electrons is investigated using the modified FTDFT calculation scheme described in Sec.~\ref{sa:dft} corresponding to a transient exotic state created by the isochoric heating with XFEL or ultrashort optical laser as the source. The samples are supposed to be at ambient conditions, before the impact of the XFEL or the optical laser pulse. This means an electron temperature of 300K for BCC Li at a solid density of 0.535 g/cm$^3$, FCC Al at a solid density of 2.70 g/cm$^3$, FCC Cu at a solid density of 8.92 g/cm$^3$, and FCC Au at a solid density of 19.3 g/cm$^3$ respectively, are used. 

First we consider the nonequilibrium electronic structure created by the XFEL, using Li, Al and Cu as illustrating examples. XFEL sources have a tunable photon energy with a pulse duration on the order of 10 fs, whereas the intensity can be as high as 10$^{17}$ W/cm$^2$ for such isochoric heating experiments~\cite{Nagler2009}. Li has a K edge of 54.7 eV~\cite{henke1993x}, which lies in the ultraviolet range accessible to free electron lasers such as FLASH, LCLS, etc. The $\textsc{L}_\textsc{I}$ edge of Al is 117.8 eV, and the $\textsc{M}_\textsc{I}$ edge of Cu is 122.5 eV, both of which are in the soft X-ray range of an XFEL. With a careful choice of the photon energy, one can create holes by predominant photoionization of 1s electrons of Li, 2s electrons of Al, or 3s electrons of Cu to their Fermi level (or chemical potential for the finite temperature case) with other transitions being of minor importance. Also, electrons excited to higher energies are found to relax to the chemical potential within a few femtoseconds~\cite{Nagler2009}.

We calculate the density of states (DOS) of such nonequilibrium transient states, as plotted in Fig.~\ref{fig:Fig1}. The black solid lines show the calculated DOS without holes, whereas the red ones correspond to the DOS in which 10 electrons were removed from the inner shell orbitals (1s for Li, 2s for Al, and 3s for Cu), leaving a hole behind on each of 10 atoms. Even though the electrons do not leave the sample which remains neutral, those atoms with a hole can be considered ``ionized'' carrying a localized positive charge. The blue lines show the DOS with 32 holes in the system. The fraction of (singly) ionized atoms, for the case of Li is, therefore, 7.8$\%$, for the red lines, and 25.0$\%$, for the blue lines. For both aluminum and copper this fraction becomes 9.2$\%$, for the red lines, and 29.6$\%$, for the blue lines. The DOS of the occupied states are shown by the shaded areas to demonstrate the positions of the holes. The positions of their corresponding chemical potentials are marked with vertical dashed color lines and their values are also labelled in the legends. The peak features of the DOS far below the chemical potential correspond to the contribution of inner shell orbitals. 

There is a noticeable shift of these atom-like features towards lower energy, due to the presence of the holes. We attribute this blue shift to the reduced screening of the core electrons, which makes the Coulomb potential that the remaining electrons feel stronger \cite{Nagler2009}. Note that the shift of the orbital energies cannot be observed by the conventional FT-DFT calculations within an impurity model, because the energy reference is ill-defined for different pseudopotentials. Moreover, The shift of the 1s orbital for Li is 22.61 eV, which is substantially smaller than the shift of 34.1 eV estimated by a separate DFT calculation for an isolated atom using the atomic code ld1.x distributed with QE. The same is observed for aluminum (a shift of 16.85 eV for the 2s level in a solid, as compared to a shift of 25.7 eV in an atomic calculation). Finally, for copper a shift of 5.26 eV is found, for the 3s level in the solid, as compared to a shift of 14.7 eV, in an atomic calculation. Therefore, the frozen core approximation should be applied with caution for such calculations. Except for Li, the position of the chemical potential rises as the number of atoms with holes increases, mainly as a consequence of the increasing number of electrons (which equals the number of holes, since the system remains neutral) that are excited to the states near chemical potential. The increase of the number of excited electrons competes with the blue shift of orbital energies, resulting in the modification of band structure.  As a result, the chemical potential first drops slightly with the creation of ten 1s holes, and then rises with 32 ionized atoms involved for Li (cf. the numbers inside the figure). 

\begin{figure}[t]
\includegraphics[width=0.41\textwidth]{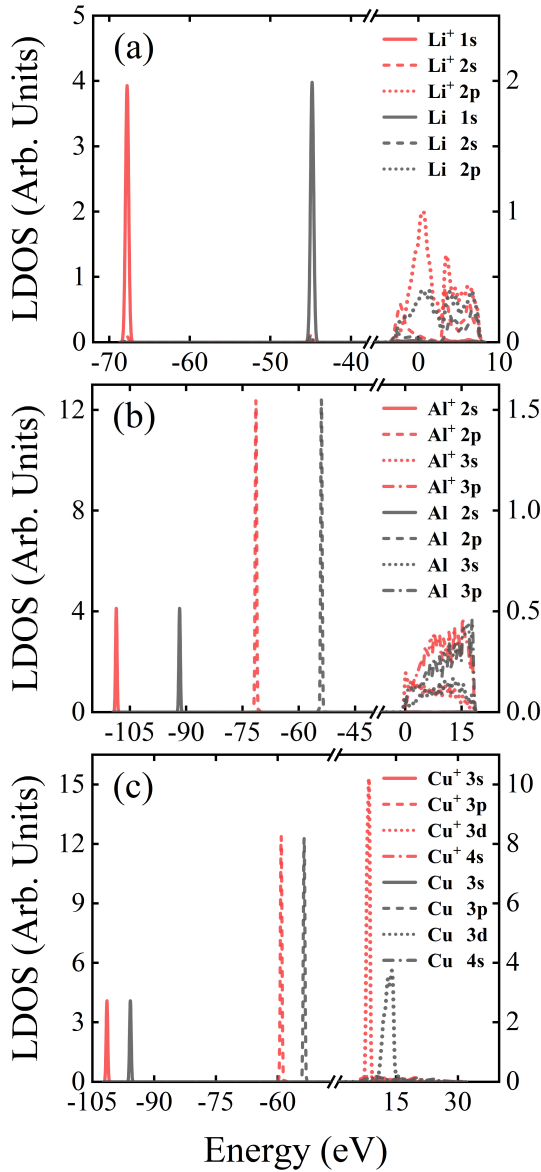}
\caption{LDOS of Li, Al and Cu with perfect lattice structure at room temperature calculated with 10 atoms having a hole in their inner shell. (a) LDOS of BCC Li of atoms with (red color) and without (black color) a hole in the 1s orbital. (b) LDOS of FCC Al of atoms with (red color) and without (black color) a hole in the 2s orbital. (c) LDOS of FCC Cu of atoms with (red color) and without (black color) a hole in the 3s orbital. The LDOS for the inner shell orbitals and for the valence band is shown with different scale (left and right axis, respectively) for better display.}
\label{fig:LDOS}
\end{figure}
Note that the position of the atomic-like line shifts towards lower energy is almost insensitive to the fraction of atoms with core holes. This also suggests that the shift has to be understood as an atomic process. We calculate the local density of states (LDOS) of Li, Al and Cu with 10 holes in the system, as shown in Fig.~\ref{fig:LDOS}, to better understand the effect of the holes on the electronic structure. The LDOS is obtained by projecting the one-electron eigenstates onto the local atomic orbitals, which indicates the contribution to the electronic structure of a specific atom. In Fig.~\ref{fig:LDOS} (a), 128 Li atoms are placed in a BCC structure with an electron temperature of 300K, among which 10 atoms have a core hole.
The atom which has a hole in its K shell is represented by red color, whereas the atom without any hole is represented by black color. Solid lines show the LDOS for the 1s orbitals, while the dashed ones show those of the 2s orbitals, and the dotted ones show those of the 2p orbitals. In Fig.~\ref{fig:LDOS} (b) [(c)], 108 Al [Cu] atoms are packed with FCC structure with electrons also at 300K, and 10 atoms having a core hole. Similar to Li, the LDOS of Al [Cu] on the atom with a hole in the 2s [3s] orbital is plotted with red color, while the LDOS on the atom without holes is plotted with black color.
Solid, dashed, dotted, and dash-dotted lines represent the LDOS for 2s [3s], 2p [3p], 3s [3d] and 3p [4s] orbitals, respectively. The shift of the 1s orbital of Li with (red solid line) or without a hole (black solid line) is clearly shown in Fig.~\ref{fig:LDOS}(a). So are the shifts of both 2s and 2p orbitals of Al, in Fig.~\ref{fig:LDOS}(b), and the shifts of both 3s and 3p orbitals of Cu, in Fig.~\ref{fig:LDOS}(c). As discussed above, this blue shift is because of the reduced screening of the core electrons. 

More interestingly, we also observe the localization of the valence band due to the reduced screening~\cite{YUAN2002275}. For Li, the LDOS of the 2s band on an atom with a 1s hole (red dashed line) has more peak structures than that on an atom without any hole (black dashed line), as shown in Fig.~\ref{fig:LDOS}~(a). The peak feature is also more significant for the M band of Al (red dotted line) with the presence of 2s hole compared to its counterpart without hole (black dotted line) in Fig.~\ref{fig:LDOS}~(b). For Cu, the appearance of a sharp peak indicates the localization of the 3d orbital (black and red dotted lines), as shown in Fig.~\ref{fig:LDOS}~(c). In fact, experiments have taken advantage of such a shift and of the localization of the valence band to circumvent the Inglis-Teller effect~\cite{inglisteller} and to study ionization potential depression (IPD)~\cite{PhysRevLett.109.065002, PhysRevLett.110.265003}. 
~\\

\begin{figure}
\includegraphics[width=0.41\textwidth]{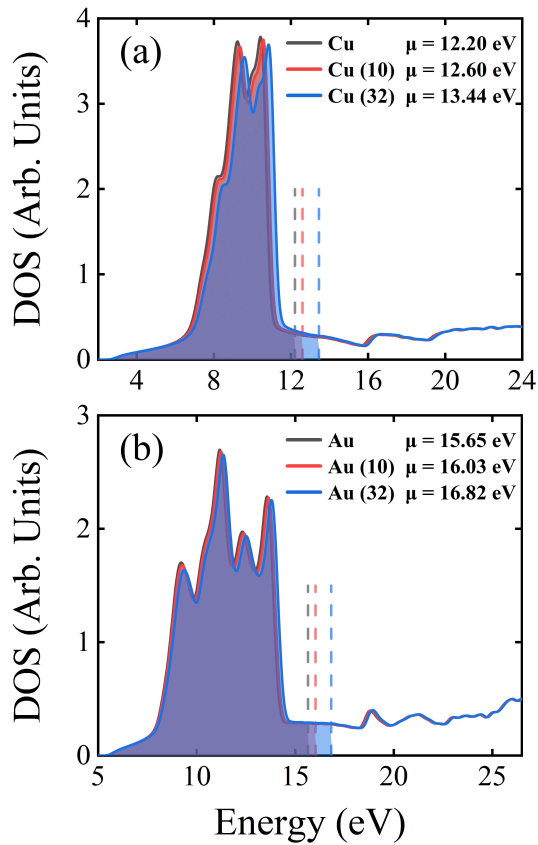}
\caption{Density of states of (a) Cu and (b) Au with perfect lattice structure at room temperature. As in Fig.~\ref{fig:Fig1}, the black solid lines show the calculated DOS, while the shaded area shows the corresponding occupied DOS. The positions of the chemical potential are marked with the vertical dashed lines and are labelled in the legend. The red [blue] lines represent the same quantity as the black lines but are calculated with 10 [32] electrons removed from their valence band (3d for Cu, and 5d for Au).
}
\label{fig:Fig3}
\end{figure}

We now investigate the electronic structure of Cu and Au with a nonequilibrium distribution of electrons, which corresponds to the transient exotic state created by the isochoric heating by an ultrashort \textit{optical laser} pulse. 
We consider a laser with a pulse duration shorter than 50 fs and an intensity of the order of 10$^{13}$ W/cm$^2$. An optical laser with 400 nm or 800 nm wavelength has a photon energy of 3.1eV or 1.55 eV, respectively. Both of them can create holes in the valence band of Cu or Au by photoexcitation. We calculate the DOS of such a non-equilibrium transient state immediately after the pulse, as plotted in Fig.~\ref{fig:Fig3}. 

The black solid lines show the calculated DOS without holes, the red ones show the DOS with 10 electrons excited from the valence band by the laser, and the blue ones show the DOS with 32 excited electrons. In this case, the electron in the valence band is no longer localized within one particular atom, therefore we can only estimate the average ionization degree of the system which is 0.09, for the red lines, and 0.30 for the blue ones. The occupied DOS are shown as the shaded areas with the same color as the DOS. The positions of their corresponding chemical potentials are marked with vertical dashed colored lines, and their values are given in the legends. This time, without the competing effect of the shift of inner shell orbitals towards lower energy, the rise of the position of chemical potential is more evident.

\begin{figure}
\includegraphics[width=0.43\textwidth]{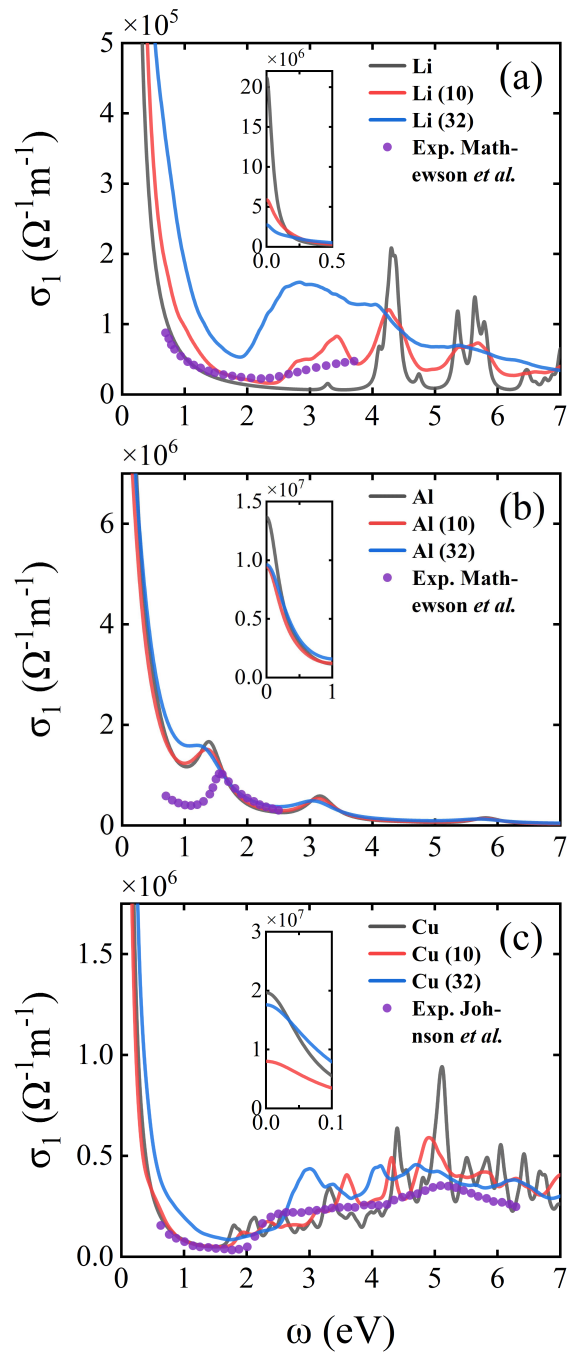}
\caption{Calculated real part of electrical conductivity of Li, Al, and Cu placed in their perfect lattice structure at room temperature with variable number of holes in their inner shell. Experimental measurements are plotted as purple dots for comparison. Inset: Zoom-in of the DC limit. (a) Li: the black line corresponds to the case without holes, while the red and blue lines represent the one with 10 and 32 atoms having an electron removed from 1s orbital, respectively. (b) The same for Al. The red and blue lines correspond to the cases of 10 and 32 atoms having an electron removed from the 2s orbital, respectively. (c) The same for Cu: The red and blue lines correspond to the cases with 10 and 32 atoms having an electron removed from the 3s orbital, respectively. Experimental data are marked with purple dots~\cite{doi:10.1080/14786437208229308, Mathewson_1971,PhysRevB.11.1315}. }
\label{fig:AC300K1}
\end{figure}

\subsection{Electrical conductivity}\label{ss:sigma}

\begin{figure}
\includegraphics[width=0.43\textwidth]{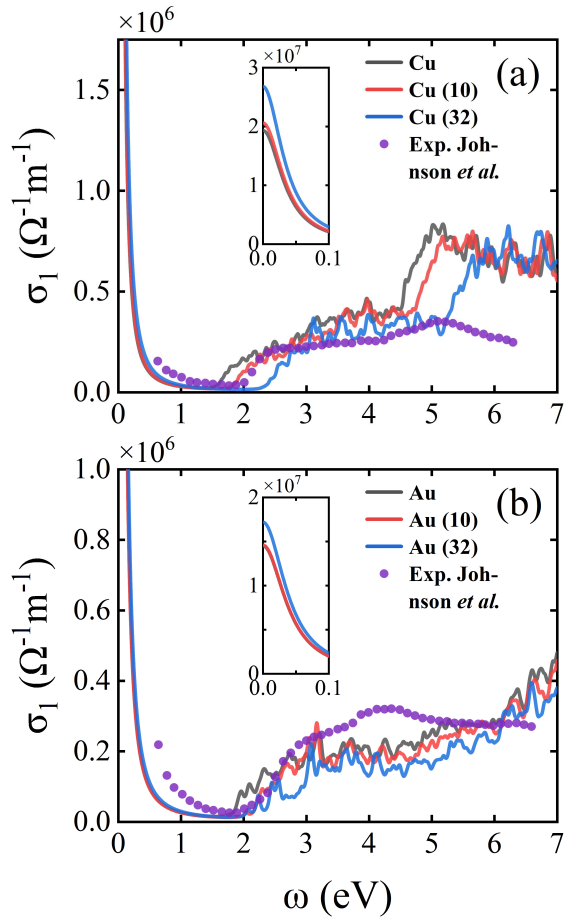}
\caption{Calculated real part of the electrical conductivity of Cu and Au placed in their perfect lattice structure at room temperature with variable number of holes in their \textit{valence band}. Experimental measurements are plotted as purple dots for comparison~\cite{PhysRevB.11.1315, PhysRevB.6.4370}. Inset: Zoom into the DC limit. (a) Cu: the black line represents the AC conductivity without holes, while the red (blue) line represents the one with 10 holes (32 holes) in the 3d orbital. (b) Au:  red (blue) line represents the case with 10 (32) holes in the 5d orbital.}
\label{fig:AC300K2}
\end{figure}

The electrical conductivity, often being linked to the thermal conductivity via the Wiedemann-Franz Law, is an important transport property of WDM~\cite{congwang2013}, as well as a useful diagnostic technique 
for the change of the electronic structure of the system~\cite{Celliers677}. Based on linear respond theory, the Kubo-Greenwood formula is widely used to study the electrical conductivity of WDM~\cite{PhysRevB.90.035121}, because the probe laser is usually weak and will not change the electronic structure significantly. The probe pulse can have a duration that is similar to that of the pump laser (or the XFEL), on the order of 10 fs, which is comparable to the estimated lifetime of the hole. We show in Fig.~\ref{fig:AC300K1} the calculated electrical conductivity of the transient exotic state of Li, Al and Cu created by the isochoric heating by an XFEL pulse corresponding to the electronic structure calculation in Sec.~\ref{sa:dos}. The black lines represent the electrical conductivity of Li, Al, and  Cu without holes, while the red and blue lines represent the one with 10 atoms and 32 atoms having a hole in their inner shell orbitals, respectively. For reference, experimental measurements for metal films at room temperature~\cite{doi:10.1080/14786437208229308, Mathewson_1971,PhysRevB.11.1315} are also shown as dots in the figure. 
Further, the width of the Dirac delta function in equation (\ref{sa:sigma1'}) and (\ref{sa:sigma1''}) is chosen as to achieve the best agreement with the experimental data. For Li, a double-peak feature appears between 2.8 eV and 3.4 eV 
due to the more localized L band with 10 holes in the system, and the peak near 2.8 eV becomes even higher for the case with 32 holes. The presence of holes in the inner shell of both Al and Cu, gives rise to a smoothening effect for the conductivity, that is attributed to the broadening of the energy levels due to different degrees of ionization of the atoms. 

The electrical conductivity near the DC limit is zoomed into in the inset. An initial drop of the DC conductivity in the case of 10 ionized atoms is observed for all these three materials. For Li, the DC conductivity further decreases when the number of ionized atoms in the system is increased to 32, while it rises for both Al and Cu. The contribution of the DC conductivity comes from the transition between valence states with energies close to each other, and the different behavior is attributed to the competition between lowering of orbital energies, due to the presence of holes, which increases the energy difference of orbitals, and the rise of the chemical potential, due to excited electrons, which means more valence states with close energies near the chemical potential. As a result, the change of the DC conductivity due to inner shell holes depends sensitively on the electronic structure of the material near its chemical potential. 

~\\

In order to show the effect of \textit{holes created in the valence band}, we also calculate the electrical conductivity of the transient exotic state of Cu and Au that is created via isochoric heating by an optical laser. The atoms remain in a FCC lattice structure with an electron temperature of 300K. The black lines in Fig.~\ref{fig:AC300K2} represent the electrical conductivity of Cu and Au without holes, while the red (blue) line corresponds to 10 (32) electrons in the valence band excited by the pump laser. We also show experimental data (dots) at room temperature for reference~\cite{PhysRevB.11.1315, PhysRevB.6.4370}. 

We first notice that the electrical conductivity starts to rise at around 1.5 eV, for Cu without holes. When 10 holes appear in the valence band, this rise shifts towards higher energies, and this shift is larger when the number of holes is increased to 32. This effect is attributed to the shift of the chemical potential because the major contribution to the rise of electrical conductivity at that energy range is the transition of 3d electrons to states around the chemical potential. A similar shift of the conductivity is  observed for Au. The rise of the chemical potential is more significant for the photoexcitation by an ultrafast laser than for photoionization by an XFEL. Thus, a stronger shift of the conductivity towards higher frequency is observed in the former case. Even though the calculated electrical conductivity with holes in the valence band shows better agreement with the experimental measurement, we cannot claim that the result is actually more accurate. The reason is that DFT is known for  notoriously underestimating the band gap. Therefore, we can only predict the shift due to the presence of holes without the exact value of such a shift. 

Finally, we also zoom into the electrical conductivity near the DC limit, see the inset. Contrary to the calculation for the case of inner shell holes, the DC conductivity increases when the number of holes in the valence band grows. Similar to what we have discussed for the DOS, this is caused by the rise of the chemical potential due to the increasing number of excited electrons. 

\subsection{Finite temperature effect}\label{ss:thermal}
\begin{figure}
\includegraphics[width=0.43\textwidth]{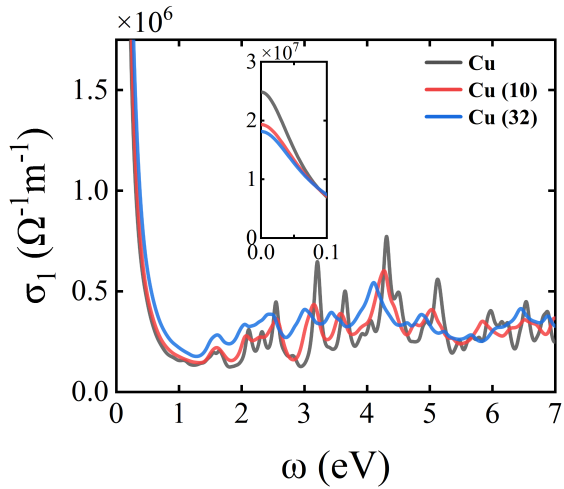}
\caption{Real part of the electrical conductivity of FCC Cu at an electron temperature of 1~eV with variable number of holes in the inner shell. Back line:  no holes; the red (blue) line: 10 (32) electron removed from the 3s orbital, resulting in 10 (32) atoms having one hole. Inset: zoom into the static limit. 
}
\label{fig:Cu1eV1}
\end{figure}
\begin{figure}
\includegraphics[width=0.43\textwidth]{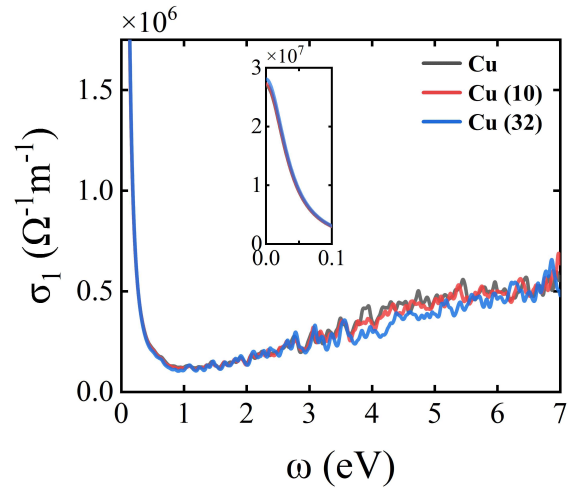}
\caption{
Same as Fig~\ref{fig:Cu1eV1}, but for the case that the holes are created in the valence band. The holes are not bounded to specific atoms in this case.}
\label{fig:Cu1eV2}
\end{figure}
In this subsection, the finite temperature effect on the nonequilibrium transient WDM states is discussed. This includes the effect of the finite temperature of the electronic subsystem, the finite temperature exchange-correlation (FTXC) interaction and the structure of the ions. Using copper as an illustrating example, we first perform the calculations of the electrical conductivity with holes in its inner shell, at an electron temperature of 1 eV, as shown in Fig.~\ref{fig:Cu1eV1}. Note that the atoms are still placed in a perfect FCC lattice structure for this case. Compared to the case of an electronic temperature of 300K, shown in Fig.~\ref{fig:AC300K1}~(c), the electrical conductivity at 1eV, with or without holes, has less structure. Particularly, in all cases it rises up between 0.5 eV and 1.5 eV, after which the curves become overall flatter. At the same time, the effect of holes in the inner shell orbital on the electrical conductivity is similar at an electronic temperature of both 300K and 1 eV. For the holes in the valence band, as shown in Fig.~\ref{fig:Cu1eV2}, however, the higher electronic temperature of 1 eV wipes out nearly all the effects caused by the holes, and only a small shift of the electrical conductivity is observed. This difference is due to the fact that the electronic temperature of 1 eV has a negligible effect on the 3s orbital which lies $\sim$70 eV below the chemical potential but considerably affects the band structure slightly below the chemical potential, which mainly consists of 3d electrons.
\begin{figure}
\includegraphics[width=0.43\textwidth]{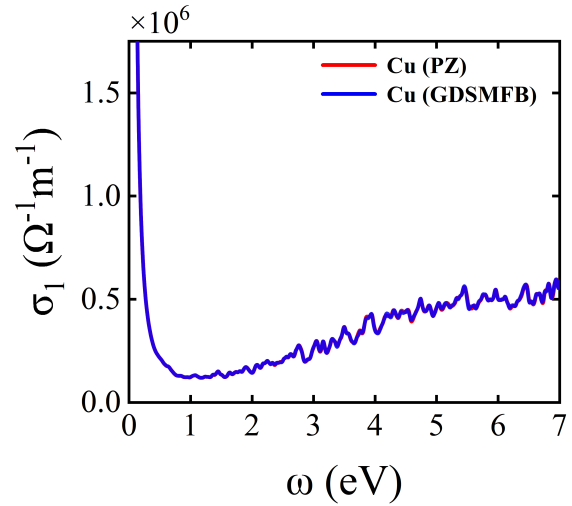}
\caption{Finite temperature effect of the exchange-correlation potential on the electrical conductivity of FCC Cu at an electronic temperature of 1~eV. Blue curve: finite temperature exchange correlation functional based on the GDSMFB parametrization~\cite{groth_prl17}. Red curve: zero-temperature functional with the PZ parametrization~\cite{PerdewZunger}. }
\label{fig:FTXC}
\end{figure}

In DFT, many-body effects of the electrons are accounted for by the exchange-correlation functional, $E_{xc}$. In particular, if the exact functional would be used, one could reproduce the exact solution of the original many-body problem of interest. For practical applications, however, this term has to be approximated. 
In practice, it turned out to be successful to solve the special case of the spatially uniform (but Coulomb interacting) electron gas (UEG) using exact quantum Monte Carlo (QMC) simulations, first provided by Ceperley and Alder~\cite{PhysRevLett.45.566}. These data were then used to construct approximate exchange-correlation functionals $E_{xc}(r_s)$ for ground state DFT simulations of real, more complicated materials. As a result, how the exchange-correlation is constructed is crucial to the accuracy of the DFT method. 

However, previous functionals, until recently, were limited to the case of zero temperature, which is adequate for many condensed matter applications, but is doomed to fail when it comes to WDM. In this case finite temperature and entropic effects in the exchange correlation functional are becoming important, and, instead of $E_{xc}$, an accurate exchange correlation free energy, $F_{xc}$, has to be provided as an input for FTDFT. Based on the recently reported \textit{ab initio} QMC data for the UEG at finite temperature~\cite{dornheim_physrep_18,schoof_prl15,dornheim_prl16}, an FTXC free energy functional has been parameterized by Groth \textit{et al.} (GDSMFB)~\cite{groth_prl17}. Here we compare the  electrical conductivity calculated with this new functional to the well-known zero-temperature functional parameterized by Perdew and Zunger~\cite{PerdewZunger}. The results are shown in Fig.~\ref{fig:FTXC} to explore the finite temperature effect on the exchange-correlation functional for the nonequilibrium transient state. Only little differences are observed between the two calculations. This is, of course, due to the relatively low temperature. We note that finite temperature effects on the DC conductivity, as high as 15\% were reported in Ref.~\cite{PhysRevE.93.063207} for cases where $\Theta\sim$ 0.5 \dots 1. 

\begin{figure}
\includegraphics[width=0.43\textwidth]{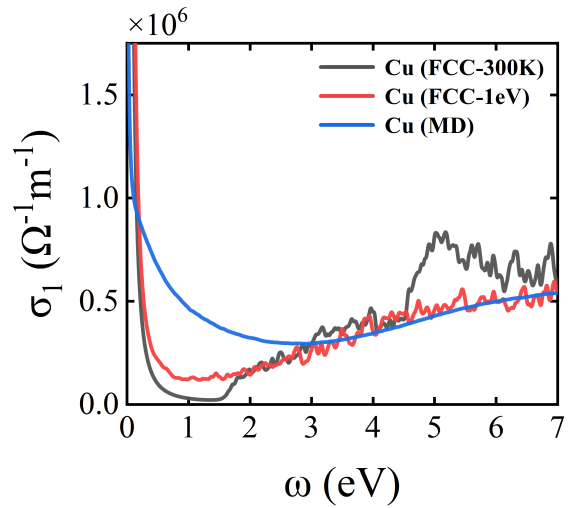}
\caption{Effect of the ionic structure on the electrical conductivity of Cu at $T=1$~eV. Red: FCC Cu at an electronic temperature of 1eV. Blue: Cu with a temperature of 1 eV for both ions and electrons. 
The error bars are smaller than the line width. The electrical conductivity of Cu at 300K (the same as the black line in Fig.~\ref{fig:AC300K2} (a) ) is also plotted as black line for reference.
}
\label{fig:MD}
\end{figure}

The effect of the ionic structure is also investigated by combining FTDFT without holes and MD simulation, where the temperature of both the ions and the electrons are set to 1 eV. The electrical conductivity, calculated by averaging over five snapshots randomly taken from the trajectory, is shown in Fig.~\ref{fig:MD}. The statistical errors  
of the real part of the electrical conductivity due to the different ionic configurations are found to be negligible (they are smaller than the line width). The liquid-like ionic structure further elevates the electrical conductivity below 3 eV compared to its increase due to electronic temperature alone. In addition, the DC conductivity decreases by an order of magnitude compared to that of the perfect FCC structure. 

\section{Conclusion} \label{s:conclusion}

Using lithium, aluminum, copper and gold as illustrating examples, we have presented a theoretical investigation to the nonequilibrium transient states generated by isochoric heating technique. We developed a modified FTDFT method by introducing a nonequilibrium distribution of electrons into the simulation. The effect of holes created by the photoexcitation of electrons in either, the inner shells or the valence band, on the electronic structure was discussed in detail. These cases can be realized by choosing either an XFEL pulse or a short laser pulse as pump source. 

The presence of holes in the inner shell results in the localization of the valence band due to the reduced screening of core electrons. In contrast, the presence of holes in the valence band mainly increases the chemical potential because of the increased number of  excited electrons. The electrical conductivity of such a nonequilibrium transient state was calculated by applying the Kubo-Greenwood formula,  with a special treatment for the DC limit. The two competing effects that were mentioned above, i.e., the lowering of orbital energies due to the presence of holes, and the rise of the chemical potential due to excited electrons, are the key factors for the change of the electrical conductivity. The hole in the valence band results in a shift of the conductivity towards higher frequency because of the dominating latter factor, while the effect of holes in the inner shell on the electrical conductivity is material-dependent due to the competition of both factors. 

We also discussed finite-temperature effects on the electrical conductivity. A finite electronic temperature results in an increase of the conductivity, for low frequency. On the other hand, the disorder in the ionic structure further appearing when the lattice is heated, also increases the conductivity. The hole in the valence band is more easily affected by the finite temperature compared to that in the inner shell, because of its relatively small energy with respect to the chemical potential. Finally the finite-temperature effect on the exchange-correlation energy is found to be small, for temperatures up to 1 eV discussed in this work. 

We also note that our work is qualitatively different from earlier calculations of optical excitation of holes in semiconductors, e.g.~\cite{kwong-etal.98pss,medvedev_10,anisimov1974electron}. In our case a significantly greater portion of excited (ionized) atoms may be involved giving rise to strong (transient) deviations of the band structure from the ground state behavior. Our results can give insight into the physical processes that can be expected in WDM experiments such as isochoric heating by means of ultrashort laser pules or free electron lasers. Of particular interest will be to extend the analysis to preheated samples that are highly ionized or in the plasma phase \cite{PhysRevLett.110.265003}. 
Then our analysis will give access to the modification of the band structure due to excitation of core electrons in dense plasmas. While we have discussed the manifestation of the nonequilibrium band structure in conductivity measurements, a more direct experimental probe should be possible via photoemission spectroscopy. Using an X-ray probe pulse with a controlled time delay a time-resolved investigation of the described nonequilibrium effects should be possible including the relaxation processes that are not included in the present study.

\section*{Acknowledgments}

The authors thank Prof. Andrew Ng for his insightful discussion about experiments. We also acknowledge discussions with Prof. Wei Kang about technical issues of FTDFT simulations. This work has been supported by the NSAF via grant No. U1830206, the National Natural Science Foundation of China via grants No. 11774429, 11874424 and 11904401 and the Science Challenge Project of China via  grant No. TZ2016001, the National Key R\&D Program of China via grant No. 2017YFA0403200, the Deutsche Forschungsgemeinschaft via grant BO1366/15, and by the HLRN via grant shp00023 for computing time. Shen Zhang acknowledges financial support from the China Scholarship Council (CSC) and from the German Academic Exchange Service (DAAD).

\bibliography{ref-sum}

\begin{thebibliography}{77}%
\makeatletter
\providecommand \@ifxundefined [1]{%
 \@ifx{#1\undefined}
}%
\providecommand \@ifnum [1]{%
 \ifnum #1\expandafter \@firstoftwo
 \else \expandafter \@secondoftwo
 \fi
}%
\providecommand \@ifx [1]{%
 \ifx #1\expandafter \@firstoftwo
 \else \expandafter \@secondoftwo
 \fi
}%
\providecommand \natexlab [1]{#1}%
\providecommand \enquote  [1]{``#1''}%
\providecommand \bibnamefont  [1]{#1}%
\providecommand \bibfnamefont [1]{#1}%
\providecommand \citenamefont [1]{#1}%
\providecommand \href@noop [0]{\@secondoftwo}%
\providecommand \href [0]{\begingroup \@sanitize@url \@href}%
\providecommand \@href[1]{\@@startlink{#1}\@@href}%
\providecommand \@@href[1]{\endgroup#1\@@endlink}%
\providecommand \@sanitize@url [0]{\catcode `\\12\catcode `\$12\catcode
  `\&12\catcode `\#12\catcode `\^12\catcode `\_12\catcode `\%12\relax}%
\providecommand \@@startlink[1]{}%
\providecommand \@@endlink[0]{}%
\providecommand \url  [0]{\begingroup\@sanitize@url \@url }%
\providecommand \@url [1]{\endgroup\@href {#1}{\urlprefix }}%
\providecommand \urlprefix  [0]{URL }%
\providecommand \Eprint [0]{\href }%
\providecommand \doibase [0]{https://doi.org/}%
\providecommand \selectlanguage [0]{\@gobble}%
\providecommand \bibinfo  [0]{\@secondoftwo}%
\providecommand \bibfield  [0]{\@secondoftwo}%
\providecommand \translation [1]{[#1]}%
\providecommand \BibitemOpen [0]{}%
\providecommand \bibitemStop [0]{}%
\providecommand \bibitemNoStop [0]{.\EOS\space}%
\providecommand \EOS [0]{\spacefactor3000\relax}%
\providecommand \BibitemShut  [1]{\csname bibitem#1\endcsname}%
\let\auto@bib@innerbib\@empty
\bibitem [{\citenamefont {Graziani}\ \emph {et~al.}(2014)\citenamefont
  {Graziani}, \citenamefont {Desjarlais}, \citenamefont {Redmer},\ and\
  \citenamefont {Trickey}}]{graziani-book}%
  \BibitemOpen
  \bibfield  {author} {\bibinfo {author} {\bibfnamefont {F.}~\bibnamefont
  {Graziani}}, \bibinfo {author} {\bibfnamefont {M.~P.}\ \bibnamefont
  {Desjarlais}}, \bibinfo {author} {\bibfnamefont {R.}~\bibnamefont {Redmer}},\
  and\ \bibinfo {author} {\bibfnamefont {S.~B.}\ \bibnamefont {Trickey}},\
  }\href@noop {} {\emph {\bibinfo {title} {Frontiers and Challenges in Warm
  Dense Matter}}}\ (\bibinfo  {publisher} {Springer},\ \bibinfo {year}
  {2014})\BibitemShut {NoStop}%
\bibitem [{\citenamefont {Bonitz}\ \emph {et~al.}(2020)\citenamefont {Bonitz},
  \citenamefont {Dornheim}, \citenamefont {Moldabekov}, \citenamefont {Zhang},
  \citenamefont {Hamann}, \citenamefont {Kählert}, \citenamefont {Filinov},
  \citenamefont {Ramakrishna},\ and\ \citenamefont
  {Vorberger}}]{bonitz_pop_20}%
  \BibitemOpen
  \bibfield  {author} {\bibinfo {author} {\bibfnamefont {M.}~\bibnamefont
  {Bonitz}}, \bibinfo {author} {\bibfnamefont {T.}~\bibnamefont {Dornheim}},
  \bibinfo {author} {\bibfnamefont {Z.~A.}\ \bibnamefont {Moldabekov}},
  \bibinfo {author} {\bibfnamefont {S.}~\bibnamefont {Zhang}}, \bibinfo
  {author} {\bibfnamefont {P.}~\bibnamefont {Hamann}}, \bibinfo {author}
  {\bibfnamefont {H.}~\bibnamefont {Kählert}}, \bibinfo {author}
  {\bibfnamefont {A.}~\bibnamefont {Filinov}}, \bibinfo {author} {\bibfnamefont
  {K.}~\bibnamefont {Ramakrishna}},\ and\ \bibinfo {author} {\bibfnamefont
  {J.}~\bibnamefont {Vorberger}},\ }\href {https://doi.org/10.1063/1.5143225}
  {\bibfield  {journal} {\bibinfo  {journal} {Phys. Plasmas}\ }\textbf
  {\bibinfo {volume} {27}},\ \bibinfo {pages} {042710} (\bibinfo {year}
  {2020})}\BibitemShut {NoStop}%
\bibitem [{\citenamefont {Bethkenhagen}\ \emph {et~al.}(2020)\citenamefont
  {Bethkenhagen}, \citenamefont {Witte}, \citenamefont {Sch\"orner},
  \citenamefont {R\"opke}, \citenamefont {D\"oppner}, \citenamefont {Kraus},
  \citenamefont {Glenzer}, \citenamefont {Sterne},\ and\ \citenamefont
  {Redmer}}]{PhysRevResearch.2.023260}%
  \BibitemOpen
  \bibfield  {author} {\bibinfo {author} {\bibfnamefont {M.}~\bibnamefont
  {Bethkenhagen}}, \bibinfo {author} {\bibfnamefont {B.~B.~L.}\ \bibnamefont
  {Witte}}, \bibinfo {author} {\bibfnamefont {M.}~\bibnamefont {Sch\"orner}},
  \bibinfo {author} {\bibfnamefont {G.}~\bibnamefont {R\"opke}}, \bibinfo
  {author} {\bibfnamefont {T.}~\bibnamefont {D\"oppner}}, \bibinfo {author}
  {\bibfnamefont {D.}~\bibnamefont {Kraus}}, \bibinfo {author} {\bibfnamefont
  {S.~H.}\ \bibnamefont {Glenzer}}, \bibinfo {author} {\bibfnamefont {P.~A.}\
  \bibnamefont {Sterne}},\ and\ \bibinfo {author} {\bibfnamefont
  {R.}~\bibnamefont {Redmer}},\ }\href
  {https://doi.org/10.1103/PhysRevResearch.2.023260} {\bibfield  {journal}
  {\bibinfo  {journal} {Phys. Rev. Research}\ }\textbf {\bibinfo {volume}
  {2}},\ \bibinfo {pages} {023260} (\bibinfo {year} {2020})}\BibitemShut
  {NoStop}%
\bibitem [{\citenamefont {Schlanges}\ \emph {et~al.}(1995)\citenamefont
  {Schlanges}, \citenamefont {Bonitz},\ and\ \citenamefont
  {Tschttschjan}}]{doi:10.1002/ctpp.2150350203}%
  \BibitemOpen
  \bibfield  {author} {\bibinfo {author} {\bibfnamefont {M.}~\bibnamefont
  {Schlanges}}, \bibinfo {author} {\bibfnamefont {M.}~\bibnamefont {Bonitz}},\
  and\ \bibinfo {author} {\bibfnamefont {A.}~\bibnamefont {Tschttschjan}},\
  }\href {https://doi.org/10.1002/ctpp.2150350203} {\bibfield  {journal}
  {\bibinfo  {journal} {Contrib. Plasma Phys.}\ }\textbf {\bibinfo {volume}
  {35}},\ \bibinfo {pages} {109} (\bibinfo {year} {1995})}\BibitemShut
  {NoStop}%
\bibitem [{\citenamefont {Zhang}\ \emph
  {et~al.}(2016{\natexlab{a}})\citenamefont {Zhang}, \citenamefont {Wang},
  \citenamefont {Kang}, \citenamefont {Zhang},\ and\ \citenamefont
  {He}}]{extFPMD}%
  \BibitemOpen
  \bibfield  {author} {\bibinfo {author} {\bibfnamefont {S.}~\bibnamefont
  {Zhang}}, \bibinfo {author} {\bibfnamefont {H.}~\bibnamefont {Wang}},
  \bibinfo {author} {\bibfnamefont {W.}~\bibnamefont {Kang}}, \bibinfo {author}
  {\bibfnamefont {P.}~\bibnamefont {Zhang}},\ and\ \bibinfo {author}
  {\bibfnamefont {X.~T.}\ \bibnamefont {He}},\ }\href
  {https://doi.org/10.1063/1.4947212} {\bibfield  {journal} {\bibinfo
  {journal} {Phys. Plasmas}\ }\textbf {\bibinfo {volume} {23}},\ \bibinfo
  {pages} {042707} (\bibinfo {year} {2016}{\natexlab{a}})}\BibitemShut
  {NoStop}%
\bibitem [{\citenamefont {Lu}\ \emph {et~al.}(2019)\citenamefont {Lu},
  \citenamefont {Kang}, \citenamefont {Wang}, \citenamefont {Gao},\ and\
  \citenamefont {Dai}}]{Lu_2019}%
  \BibitemOpen
  \bibfield  {author} {\bibinfo {author} {\bibfnamefont {B.}~\bibnamefont
  {Lu}}, \bibinfo {author} {\bibfnamefont {D.}~\bibnamefont {Kang}}, \bibinfo
  {author} {\bibfnamefont {D.}~\bibnamefont {Wang}}, \bibinfo {author}
  {\bibfnamefont {T.}~\bibnamefont {Gao}},\ and\ \bibinfo {author}
  {\bibfnamefont {J.}~\bibnamefont {Dai}},\ }\href
  {https://doi.org/10.1088/0256-307x/36/10/103102} {\bibfield  {journal}
  {\bibinfo  {journal} {Chin. Phys. Lett.}\ }\textbf {\bibinfo {volume} {36}},\
  \bibinfo {pages} {103102} (\bibinfo {year} {2019})}\BibitemShut {NoStop}%
\bibitem [{\citenamefont {Hu}\ \emph {et~al.}(2014)\citenamefont {Hu},
  \citenamefont {Collins}, \citenamefont {Boehly}, \citenamefont {Kress},
  \citenamefont {Goncharov},\ and\ \citenamefont
  {Skupsky}}]{PhysRevE.89.043105}%
  \BibitemOpen
  \bibfield  {author} {\bibinfo {author} {\bibfnamefont {S.~X.}\ \bibnamefont
  {Hu}}, \bibinfo {author} {\bibfnamefont {L.~A.}\ \bibnamefont {Collins}},
  \bibinfo {author} {\bibfnamefont {T.~R.}\ \bibnamefont {Boehly}}, \bibinfo
  {author} {\bibfnamefont {J.~D.}\ \bibnamefont {Kress}}, \bibinfo {author}
  {\bibfnamefont {V.~N.}\ \bibnamefont {Goncharov}},\ and\ \bibinfo {author}
  {\bibfnamefont {S.}~\bibnamefont {Skupsky}},\ }\href
  {https://doi.org/10.1103/PhysRevE.89.043105} {\bibfield  {journal} {\bibinfo
  {journal} {Phys. Rev. E}\ }\textbf {\bibinfo {volume} {89}},\ \bibinfo
  {pages} {043105} (\bibinfo {year} {2014})}\BibitemShut {NoStop}%
\bibitem [{\citenamefont {Kang}\ \emph {et~al.}(2020)\citenamefont {Kang},
  \citenamefont {Hou}, \citenamefont {Zeng},\ and\ \citenamefont
  {Dai}}]{doi:10.1063/5.0008231}%
  \BibitemOpen
  \bibfield  {author} {\bibinfo {author} {\bibfnamefont {D.}~\bibnamefont
  {Kang}}, \bibinfo {author} {\bibfnamefont {Y.}~\bibnamefont {Hou}}, \bibinfo
  {author} {\bibfnamefont {Q.}~\bibnamefont {Zeng}},\ and\ \bibinfo {author}
  {\bibfnamefont {J.}~\bibnamefont {Dai}},\ }\href
  {https://doi.org/10.1063/5.0008231} {\bibfield  {journal} {\bibinfo
  {journal} {Matter Radiat. at Extremes}\ }\textbf {\bibinfo {volume} {5}},\
  \bibinfo {pages} {055401} (\bibinfo {year} {2020})}\BibitemShut {NoStop}%
\bibitem [{\citenamefont {Falk}(2018)}]{falk_2018}%
  \BibitemOpen
  \bibfield  {author} {\bibinfo {author} {\bibfnamefont {K.}~\bibnamefont
  {Falk}},\ }\href {https://doi.org/10.1017/hpl.2018.53} {\bibfield  {journal}
  {\bibinfo  {journal} {High Power Laser Sci. Eng.}\ }\textbf {\bibinfo
  {volume} {6}},\ \bibinfo {pages} {e59} (\bibinfo {year} {2018})}\BibitemShut
  {NoStop}%
\bibitem [{\citenamefont {Saemann}\ \emph {et~al.}(1999)\citenamefont
  {Saemann}, \citenamefont {Eidmann}, \citenamefont {Golovkin}, \citenamefont
  {Mancini}, \citenamefont {Andersson}, \citenamefont {F\"orster},\ and\
  \citenamefont {Witte}}]{PhysRevLett.82.4843}%
  \BibitemOpen
  \bibfield  {author} {\bibinfo {author} {\bibfnamefont {A.}~\bibnamefont
  {Saemann}}, \bibinfo {author} {\bibfnamefont {K.}~\bibnamefont {Eidmann}},
  \bibinfo {author} {\bibfnamefont {I.~E.}\ \bibnamefont {Golovkin}}, \bibinfo
  {author} {\bibfnamefont {R.~C.}\ \bibnamefont {Mancini}}, \bibinfo {author}
  {\bibfnamefont {E.}~\bibnamefont {Andersson}}, \bibinfo {author}
  {\bibfnamefont {E.}~\bibnamefont {F\"orster}},\ and\ \bibinfo {author}
  {\bibfnamefont {K.}~\bibnamefont {Witte}},\ }\href
  {https://doi.org/10.1103/PhysRevLett.82.4843} {\bibfield  {journal} {\bibinfo
   {journal} {Phys. Rev. Lett.}\ }\textbf {\bibinfo {volume} {82}},\ \bibinfo
  {pages} {4843} (\bibinfo {year} {1999})}\BibitemShut {NoStop}%
\bibitem [{\citenamefont {Chen}\ \emph {et~al.}(2013)\citenamefont {Chen},
  \citenamefont {Holst}, \citenamefont {Kirkwood}, \citenamefont {Sametoglu},
  \citenamefont {Reid}, \citenamefont {Tsui}, \citenamefont {Recoules},\ and\
  \citenamefont {Ng}}]{PhysRevLett.110.135001}%
  \BibitemOpen
  \bibfield  {author} {\bibinfo {author} {\bibfnamefont {Z.}~\bibnamefont
  {Chen}}, \bibinfo {author} {\bibfnamefont {B.}~\bibnamefont {Holst}},
  \bibinfo {author} {\bibfnamefont {S.~E.}\ \bibnamefont {Kirkwood}}, \bibinfo
  {author} {\bibfnamefont {V.}~\bibnamefont {Sametoglu}}, \bibinfo {author}
  {\bibfnamefont {M.}~\bibnamefont {Reid}}, \bibinfo {author} {\bibfnamefont
  {Y.~Y.}\ \bibnamefont {Tsui}}, \bibinfo {author} {\bibfnamefont
  {V.}~\bibnamefont {Recoules}},\ and\ \bibinfo {author} {\bibfnamefont
  {A.}~\bibnamefont {Ng}},\ }\href
  {https://doi.org/10.1103/PhysRevLett.110.135001} {\bibfield  {journal}
  {\bibinfo  {journal} {Phys. Rev. Lett.}\ }\textbf {\bibinfo {volume} {110}},\
  \bibinfo {pages} {135001} (\bibinfo {year} {2013})}\BibitemShut {NoStop}%
\bibitem [{\citenamefont {Sentoku}\ \emph {et~al.}(2007)\citenamefont
  {Sentoku}, \citenamefont {Kemp}, \citenamefont {Presura}, \citenamefont
  {Bakeman},\ and\ \citenamefont {Cowan}}]{doi:10.1063/1.2816439}%
  \BibitemOpen
  \bibfield  {author} {\bibinfo {author} {\bibfnamefont {Y.}~\bibnamefont
  {Sentoku}}, \bibinfo {author} {\bibfnamefont {A.~J.}\ \bibnamefont {Kemp}},
  \bibinfo {author} {\bibfnamefont {R.}~\bibnamefont {Presura}}, \bibinfo
  {author} {\bibfnamefont {M.~S.}\ \bibnamefont {Bakeman}},\ and\ \bibinfo
  {author} {\bibfnamefont {T.~E.}\ \bibnamefont {Cowan}},\ }\href
  {https://doi.org/10.1063/1.2816439} {\bibfield  {journal} {\bibinfo
  {journal} {Phys. Plasmas}\ }\textbf {\bibinfo {volume} {14}},\ \bibinfo
  {pages} {122701} (\bibinfo {year} {2007})}\BibitemShut {NoStop}%
\bibitem [{\citenamefont {Patel}\ \emph {et~al.}(2003)\citenamefont {Patel},
  \citenamefont {Mackinnon}, \citenamefont {Key}, \citenamefont {Cowan},
  \citenamefont {Foord}, \citenamefont {Allen}, \citenamefont {Price},
  \citenamefont {Ruhl}, \citenamefont {Springer},\ and\ \citenamefont
  {Stephens}}]{PhysRevLett.91.125004}%
  \BibitemOpen
  \bibfield  {author} {\bibinfo {author} {\bibfnamefont {P.~K.}\ \bibnamefont
  {Patel}}, \bibinfo {author} {\bibfnamefont {A.~J.}\ \bibnamefont
  {Mackinnon}}, \bibinfo {author} {\bibfnamefont {M.~H.}\ \bibnamefont {Key}},
  \bibinfo {author} {\bibfnamefont {T.~E.}\ \bibnamefont {Cowan}}, \bibinfo
  {author} {\bibfnamefont {M.~E.}\ \bibnamefont {Foord}}, \bibinfo {author}
  {\bibfnamefont {M.}~\bibnamefont {Allen}}, \bibinfo {author} {\bibfnamefont
  {D.~F.}\ \bibnamefont {Price}}, \bibinfo {author} {\bibfnamefont
  {H.}~\bibnamefont {Ruhl}}, \bibinfo {author} {\bibfnamefont {P.~T.}\
  \bibnamefont {Springer}},\ and\ \bibinfo {author} {\bibfnamefont
  {R.}~\bibnamefont {Stephens}},\ }\href
  {https://doi.org/10.1103/PhysRevLett.91.125004} {\bibfield  {journal}
  {\bibinfo  {journal} {Phys. Rev. Lett.}\ }\textbf {\bibinfo {volume} {91}},\
  \bibinfo {pages} {125004} (\bibinfo {year} {2003})}\BibitemShut {NoStop}%
\bibitem [{\citenamefont {Roth}\ \emph {et~al.}(2009)\citenamefont {Roth},
  \citenamefont {Alber}, \citenamefont {Bagnoud}, \citenamefont {Brown},
  \citenamefont {Clarke}, \citenamefont {Daido}, \citenamefont {Fernandez},
  \citenamefont {Flippo}, \citenamefont {Gaillard}, \citenamefont {Gauthier}
  \emph {et~al.}}]{Roth_2009}%
  \BibitemOpen
  \bibfield  {author} {\bibinfo {author} {\bibfnamefont {M.}~\bibnamefont
  {Roth}}, \bibinfo {author} {\bibfnamefont {I.}~\bibnamefont {Alber}},
  \bibinfo {author} {\bibfnamefont {V.}~\bibnamefont {Bagnoud}}, \bibinfo
  {author} {\bibfnamefont {C.~R.~D.}\ \bibnamefont {Brown}}, \bibinfo {author}
  {\bibfnamefont {R.}~\bibnamefont {Clarke}}, \bibinfo {author} {\bibfnamefont
  {H.}~\bibnamefont {Daido}}, \bibinfo {author} {\bibfnamefont
  {J.}~\bibnamefont {Fernandez}}, \bibinfo {author} {\bibfnamefont
  {K.}~\bibnamefont {Flippo}}, \bibinfo {author} {\bibfnamefont
  {S.}~\bibnamefont {Gaillard}}, \bibinfo {author} {\bibfnamefont
  {C.}~\bibnamefont {Gauthier}}, \emph {et~al.},\ }\href
  {https://doi.org/10.1088/0741-3335/51/12/124039} {\bibfield  {journal}
  {\bibinfo  {journal} {Plasma Phys. Control. Fusion}\ }\textbf {\bibinfo
  {volume} {51}},\ \bibinfo {pages} {124039} (\bibinfo {year}
  {2009})}\BibitemShut {NoStop}%
\bibitem [{\citenamefont {Mancic}\ \emph {et~al.}(2010)\citenamefont {Mancic},
  \citenamefont {Robiche}, \citenamefont {Antici}, \citenamefont {Audebert},
  \citenamefont {Blancard}, \citenamefont {Combis}, \citenamefont {Dorchies},
  \citenamefont {Faussurier}, \citenamefont {Fourmaux}, \citenamefont
  {Harmand}, \citenamefont {Kodama}, \citenamefont {Lancia}, \citenamefont
  {Mazevet}, \citenamefont {Nakatsutsumi}, \citenamefont {Peyrusse},
  \citenamefont {Recoules}, \citenamefont {Renaudin}, \citenamefont
  {Shepherd},\ and\ \citenamefont {Fuchs}}]{MANCIC201021}%
  \BibitemOpen
  \bibfield  {author} {\bibinfo {author} {\bibfnamefont {A.}~\bibnamefont
  {Mancic}}, \bibinfo {author} {\bibfnamefont {J.}~\bibnamefont {Robiche}},
  \bibinfo {author} {\bibfnamefont {P.}~\bibnamefont {Antici}}, \bibinfo
  {author} {\bibfnamefont {P.}~\bibnamefont {Audebert}}, \bibinfo {author}
  {\bibfnamefont {C.}~\bibnamefont {Blancard}}, \bibinfo {author}
  {\bibfnamefont {P.}~\bibnamefont {Combis}}, \bibinfo {author} {\bibfnamefont
  {F.}~\bibnamefont {Dorchies}}, \bibinfo {author} {\bibfnamefont
  {G.}~\bibnamefont {Faussurier}}, \bibinfo {author} {\bibfnamefont
  {S.}~\bibnamefont {Fourmaux}}, \bibinfo {author} {\bibfnamefont
  {M.}~\bibnamefont {Harmand}}, \bibinfo {author} {\bibfnamefont
  {R.}~\bibnamefont {Kodama}}, \bibinfo {author} {\bibfnamefont
  {L.}~\bibnamefont {Lancia}}, \bibinfo {author} {\bibfnamefont
  {S.}~\bibnamefont {Mazevet}}, \bibinfo {author} {\bibfnamefont
  {M.}~\bibnamefont {Nakatsutsumi}}, \bibinfo {author} {\bibfnamefont
  {O.}~\bibnamefont {Peyrusse}}, \bibinfo {author} {\bibfnamefont
  {V.}~\bibnamefont {Recoules}}, \bibinfo {author} {\bibfnamefont
  {P.}~\bibnamefont {Renaudin}}, \bibinfo {author} {\bibfnamefont
  {R.}~\bibnamefont {Shepherd}},\ and\ \bibinfo {author} {\bibfnamefont
  {J.}~\bibnamefont {Fuchs}},\ }\href
  {https://doi.org/https://doi.org/10.1016/j.hedp.2009.06.008} {\bibfield
  {journal} {\bibinfo  {journal} {High Energy Density Phys.}\ }\textbf
  {\bibinfo {volume} {6}},\ \bibinfo {pages} {21 } (\bibinfo {year}
  {2010})}\BibitemShut {NoStop}%
\bibitem [{\citenamefont {Vinko}\ \emph {et~al.}(2012)\citenamefont {Vinko},
  \citenamefont {Ciricosta}, \citenamefont {Cho}, \citenamefont {Engelhorn},
  \citenamefont {Chung}, \citenamefont {Brown}, \citenamefont {Burian},
  \citenamefont {Chalupský}, \citenamefont {Falcone}, \citenamefont {Graves}
  \emph {et~al.}}]{Vinko2012}%
  \BibitemOpen
  \bibfield  {author} {\bibinfo {author} {\bibfnamefont {S.~M.}\ \bibnamefont
  {Vinko}}, \bibinfo {author} {\bibfnamefont {O.}~\bibnamefont {Ciricosta}},
  \bibinfo {author} {\bibfnamefont {B.~I.}\ \bibnamefont {Cho}}, \bibinfo
  {author} {\bibfnamefont {K.}~\bibnamefont {Engelhorn}}, \bibinfo {author}
  {\bibfnamefont {H.~K.}\ \bibnamefont {Chung}}, \bibinfo {author}
  {\bibfnamefont {C.~R.~D.}\ \bibnamefont {Brown}}, \bibinfo {author}
  {\bibfnamefont {T.}~\bibnamefont {Burian}}, \bibinfo {author} {\bibfnamefont
  {J.}~\bibnamefont {Chalupský}}, \bibinfo {author} {\bibfnamefont {R.~W.}\
  \bibnamefont {Falcone}}, \bibinfo {author} {\bibfnamefont {C.}~\bibnamefont
  {Graves}}, \emph {et~al.},\ }\href {https://doi.org/10.1038/nature10746}
  {\bibfield  {journal} {\bibinfo  {journal} {Nature}\ }\textbf {\bibinfo
  {volume} {482}},\ \bibinfo {pages} {59} (\bibinfo {year} {2012})}\BibitemShut
  {NoStop}%
\bibitem [{\citenamefont {Hau-Riege}\ \emph {et~al.}(2012)\citenamefont
  {Hau-Riege}, \citenamefont {Graf}, \citenamefont {D\"oppner}, \citenamefont
  {London}, \citenamefont {Krzywinski}, \citenamefont {Fortmann}, \citenamefont
  {Glenzer}, \citenamefont {Frank}, \citenamefont {Sokolowski-Tinten},
  \citenamefont {Messerschmidt} \emph {et~al.}}]{xfel_carbon}%
  \BibitemOpen
  \bibfield  {author} {\bibinfo {author} {\bibfnamefont {S.~P.}\ \bibnamefont
  {Hau-Riege}}, \bibinfo {author} {\bibfnamefont {A.}~\bibnamefont {Graf}},
  \bibinfo {author} {\bibfnamefont {T.}~\bibnamefont {D\"oppner}}, \bibinfo
  {author} {\bibfnamefont {R.~A.}\ \bibnamefont {London}}, \bibinfo {author}
  {\bibfnamefont {J.}~\bibnamefont {Krzywinski}}, \bibinfo {author}
  {\bibfnamefont {C.}~\bibnamefont {Fortmann}}, \bibinfo {author}
  {\bibfnamefont {S.~H.}\ \bibnamefont {Glenzer}}, \bibinfo {author}
  {\bibfnamefont {M.}~\bibnamefont {Frank}}, \bibinfo {author} {\bibfnamefont
  {K.}~\bibnamefont {Sokolowski-Tinten}}, \bibinfo {author} {\bibfnamefont
  {M.}~\bibnamefont {Messerschmidt}}, \emph {et~al.},\ }\href
  {https://doi.org/10.1103/PhysRevLett.108.217402} {\bibfield  {journal}
  {\bibinfo  {journal} {Phys. Rev. Lett.}\ }\textbf {\bibinfo {volume} {108}},\
  \bibinfo {pages} {217402} (\bibinfo {year} {2012})}\BibitemShut {NoStop}%
\bibitem [{\citenamefont {Kraus}\ \emph {et~al.}(2018)\citenamefont {Kraus},
  \citenamefont {Bachmann}, \citenamefont {Barbrel}, \citenamefont {Falcone},
  \citenamefont {Fletcher}, \citenamefont {Frydrych}, \citenamefont {Gamboa},
  \citenamefont {Gauthier}, \citenamefont {Gericke}, \citenamefont {Glenzer},
  \citenamefont {Göde}, \citenamefont {Granados}, \citenamefont {Hartley},
  \citenamefont {Helfrich}, \citenamefont {Lee}, \citenamefont {Nagler},
  \citenamefont {Ravasio}, \citenamefont {Schumaker}, \citenamefont
  {Vorberger},\ and\ \citenamefont {Döppner}}]{Kraus_2018}%
  \BibitemOpen
  \bibfield  {author} {\bibinfo {author} {\bibfnamefont {D.}~\bibnamefont
  {Kraus}}, \bibinfo {author} {\bibfnamefont {B.}~\bibnamefont {Bachmann}},
  \bibinfo {author} {\bibfnamefont {B.}~\bibnamefont {Barbrel}}, \bibinfo
  {author} {\bibfnamefont {R.~W.}\ \bibnamefont {Falcone}}, \bibinfo {author}
  {\bibfnamefont {L.~B.}\ \bibnamefont {Fletcher}}, \bibinfo {author}
  {\bibfnamefont {S.}~\bibnamefont {Frydrych}}, \bibinfo {author}
  {\bibfnamefont {E.~J.}\ \bibnamefont {Gamboa}}, \bibinfo {author}
  {\bibfnamefont {M.}~\bibnamefont {Gauthier}}, \bibinfo {author}
  {\bibfnamefont {D.~O.}\ \bibnamefont {Gericke}}, \bibinfo {author}
  {\bibfnamefont {S.~H.}\ \bibnamefont {Glenzer}}, \bibinfo {author}
  {\bibfnamefont {S.}~\bibnamefont {Göde}}, \bibinfo {author} {\bibfnamefont
  {E.}~\bibnamefont {Granados}}, \bibinfo {author} {\bibfnamefont {N.~J.}\
  \bibnamefont {Hartley}}, \bibinfo {author} {\bibfnamefont {J.}~\bibnamefont
  {Helfrich}}, \bibinfo {author} {\bibfnamefont {H.~J.}\ \bibnamefont {Lee}},
  \bibinfo {author} {\bibfnamefont {B.}~\bibnamefont {Nagler}}, \bibinfo
  {author} {\bibfnamefont {A.}~\bibnamefont {Ravasio}}, \bibinfo {author}
  {\bibfnamefont {W.}~\bibnamefont {Schumaker}}, \bibinfo {author}
  {\bibfnamefont {J.}~\bibnamefont {Vorberger}},\ and\ \bibinfo {author}
  {\bibfnamefont {T.}~\bibnamefont {Döppner}},\ }\href
  {https://doi.org/10.1088/1361-6587/aadd6c} {\bibfield  {journal} {\bibinfo
  {journal} {Plasma Phys. Control. Fusion}\ }\textbf {\bibinfo {volume} {61}},\
  \bibinfo {pages} {014015} (\bibinfo {year} {2018})}\BibitemShut {NoStop}%
\bibitem [{\citenamefont {Chen}\ \emph {et~al.}(2018)\citenamefont {Chen},
  \citenamefont {Mo}, \citenamefont {Soulard}, \citenamefont {Recoules},
  \citenamefont {Hering}, \citenamefont {Tsui}, \citenamefont {Glenzer},\ and\
  \citenamefont {Ng}}]{chen_ng}%
  \BibitemOpen
  \bibfield  {author} {\bibinfo {author} {\bibfnamefont {Z.}~\bibnamefont
  {Chen}}, \bibinfo {author} {\bibfnamefont {M.}~\bibnamefont {Mo}}, \bibinfo
  {author} {\bibfnamefont {L.}~\bibnamefont {Soulard}}, \bibinfo {author}
  {\bibfnamefont {V.}~\bibnamefont {Recoules}}, \bibinfo {author}
  {\bibfnamefont {P.}~\bibnamefont {Hering}}, \bibinfo {author} {\bibfnamefont
  {Y.~Y.}\ \bibnamefont {Tsui}}, \bibinfo {author} {\bibfnamefont {S.~H.}\
  \bibnamefont {Glenzer}},\ and\ \bibinfo {author} {\bibfnamefont
  {A.}~\bibnamefont {Ng}},\ }\href
  {https://doi.org/10.1103/PhysRevLett.121.075002} {\bibfield  {journal}
  {\bibinfo  {journal} {Phys. Rev. Lett.}\ }\textbf {\bibinfo {volume} {121}},\
  \bibinfo {pages} {075002} (\bibinfo {year} {2018})}\BibitemShut {NoStop}%
\bibitem [{\citenamefont {Zhang}\ \emph {et~al.}(2019)\citenamefont {Zhang},
  \citenamefont {Gan}, \citenamefont {Zhuo}, \citenamefont {Qiao},
  \citenamefont {Ma},\ and\ \citenamefont {Dai}}]{Hengyu}%
  \BibitemOpen
  \bibfield  {author} {\bibinfo {author} {\bibfnamefont {H.-Y.}\ \bibnamefont
  {Zhang}}, \bibinfo {author} {\bibfnamefont {L.-F.}\ \bibnamefont {Gan}},
  \bibinfo {author} {\bibfnamefont {H.-B.}\ \bibnamefont {Zhuo}}, \bibinfo
  {author} {\bibfnamefont {B.}~\bibnamefont {Qiao}}, \bibinfo {author}
  {\bibfnamefont {Y.-Y.}\ \bibnamefont {Ma}},\ and\ \bibinfo {author}
  {\bibfnamefont {J.-Y.}\ \bibnamefont {Dai}},\ }\href
  {https://doi.org/10.1103/PhysRevA.100.022122} {\bibfield  {journal} {\bibinfo
   {journal} {Phys. Rev. A}\ }\textbf {\bibinfo {volume} {100}},\ \bibinfo
  {pages} {022122} (\bibinfo {year} {2019})}\BibitemShut {NoStop}%
\bibitem [{\citenamefont {Di~Piazza}\ \emph {et~al.}(2012)\citenamefont
  {Di~Piazza}, \citenamefont {M\"uller}, \citenamefont {Hatsagortsyan},\ and\
  \citenamefont {Keitel}}]{RevModPhys.84.1177}%
  \BibitemOpen
  \bibfield  {author} {\bibinfo {author} {\bibfnamefont {A.}~\bibnamefont
  {Di~Piazza}}, \bibinfo {author} {\bibfnamefont {C.}~\bibnamefont {M\"uller}},
  \bibinfo {author} {\bibfnamefont {K.~Z.}\ \bibnamefont {Hatsagortsyan}},\
  and\ \bibinfo {author} {\bibfnamefont {C.~H.}\ \bibnamefont {Keitel}},\
  }\href {https://doi.org/10.1103/RevModPhys.84.1177} {\bibfield  {journal}
  {\bibinfo  {journal} {Rev. Mod. Phys.}\ }\textbf {\bibinfo {volume} {84}},\
  \bibinfo {pages} {1177} (\bibinfo {year} {2012})}\BibitemShut {NoStop}%
\bibitem [{\citenamefont {Kiriyama}\ \emph {et~al.}(2018)\citenamefont
  {Kiriyama}, \citenamefont {Pirozhkov}, \citenamefont {Nishiuchi},
  \citenamefont {Fukuda}, \citenamefont {Ogura}, \citenamefont {Sagisaka},
  \citenamefont {Miyasaka}, \citenamefont {Mori}, \citenamefont {Sakaki},
  \citenamefont {Dover}, \citenamefont {Kondo}, \citenamefont {Koga},
  \citenamefont {Esirkepov}, \citenamefont {Kando},\ and\ \citenamefont
  {Kondo}}]{Kiriyama}%
  \BibitemOpen
  \bibfield  {author} {\bibinfo {author} {\bibfnamefont {H.}~\bibnamefont
  {Kiriyama}}, \bibinfo {author} {\bibfnamefont {A.~S.}\ \bibnamefont
  {Pirozhkov}}, \bibinfo {author} {\bibfnamefont {M.}~\bibnamefont
  {Nishiuchi}}, \bibinfo {author} {\bibfnamefont {Y.}~\bibnamefont {Fukuda}},
  \bibinfo {author} {\bibfnamefont {K.}~\bibnamefont {Ogura}}, \bibinfo
  {author} {\bibfnamefont {A.}~\bibnamefont {Sagisaka}}, \bibinfo {author}
  {\bibfnamefont {Y.}~\bibnamefont {Miyasaka}}, \bibinfo {author}
  {\bibfnamefont {M.}~\bibnamefont {Mori}}, \bibinfo {author} {\bibfnamefont
  {H.}~\bibnamefont {Sakaki}}, \bibinfo {author} {\bibfnamefont {N.~P.}\
  \bibnamefont {Dover}}, \bibinfo {author} {\bibfnamefont {K.}~\bibnamefont
  {Kondo}}, \bibinfo {author} {\bibfnamefont {J.~K.}\ \bibnamefont {Koga}},
  \bibinfo {author} {\bibfnamefont {T.~Z.}\ \bibnamefont {Esirkepov}}, \bibinfo
  {author} {\bibfnamefont {M.}~\bibnamefont {Kando}},\ and\ \bibinfo {author}
  {\bibfnamefont {K.}~\bibnamefont {Kondo}},\ }\href
  {https://doi.org/10.1364/OL.43.002595} {\bibfield  {journal} {\bibinfo
  {journal} {Opt. Lett.}\ }\textbf {\bibinfo {volume} {43}},\ \bibinfo {pages}
  {2595} (\bibinfo {year} {2018})}\BibitemShut {NoStop}%
\bibitem [{\citenamefont {Mimura}\ \emph {et~al.}(2014)\citenamefont {Mimura},
  \citenamefont {Yumoto}, \citenamefont {Matsuyama}, \citenamefont {Koyama},
  \citenamefont {Tono}, \citenamefont {Inubushi}, \citenamefont {Togashi},
  \citenamefont {Sato}, \citenamefont {Kim}, \citenamefont {Fukui} \emph
  {et~al.}}]{mimura2014generation}%
  \BibitemOpen
  \bibfield  {author} {\bibinfo {author} {\bibfnamefont {H.}~\bibnamefont
  {Mimura}}, \bibinfo {author} {\bibfnamefont {H.}~\bibnamefont {Yumoto}},
  \bibinfo {author} {\bibfnamefont {S.}~\bibnamefont {Matsuyama}}, \bibinfo
  {author} {\bibfnamefont {T.}~\bibnamefont {Koyama}}, \bibinfo {author}
  {\bibfnamefont {K.}~\bibnamefont {Tono}}, \bibinfo {author} {\bibfnamefont
  {Y.}~\bibnamefont {Inubushi}}, \bibinfo {author} {\bibfnamefont
  {T.}~\bibnamefont {Togashi}}, \bibinfo {author} {\bibfnamefont
  {T.}~\bibnamefont {Sato}}, \bibinfo {author} {\bibfnamefont {J.}~\bibnamefont
  {Kim}}, \bibinfo {author} {\bibfnamefont {R.}~\bibnamefont {Fukui}}, \emph
  {et~al.},\ }\href {https://doi.org/10.1038/ncomms4539} {\bibfield  {journal}
  {\bibinfo  {journal} {Nat. Commun.}\ }\textbf {\bibinfo {volume} {5}},\
  \bibinfo {pages} {3539} (\bibinfo {year} {2014})}\BibitemShut {NoStop}%
\bibitem [{\citenamefont {Nagler}\ \emph {et~al.}(2009)\citenamefont {Nagler},
  \citenamefont {Zastrau}, \citenamefont {F{\"a}ustlin}, \citenamefont {Vinko},
  \citenamefont {Whitcher}, \citenamefont {Nelson}, \citenamefont
  {Sobierajski}, \citenamefont {Krzywinski}, \citenamefont {Chalupsky},
  \citenamefont {Abreu} \emph {et~al.}}]{Nagler2009}%
  \BibitemOpen
  \bibfield  {author} {\bibinfo {author} {\bibfnamefont {B.}~\bibnamefont
  {Nagler}}, \bibinfo {author} {\bibfnamefont {U.}~\bibnamefont {Zastrau}},
  \bibinfo {author} {\bibfnamefont {R.~R.}\ \bibnamefont {F{\"a}ustlin}},
  \bibinfo {author} {\bibfnamefont {S.~M.}\ \bibnamefont {Vinko}}, \bibinfo
  {author} {\bibfnamefont {T.}~\bibnamefont {Whitcher}}, \bibinfo {author}
  {\bibfnamefont {A.~J.}\ \bibnamefont {Nelson}}, \bibinfo {author}
  {\bibfnamefont {R.}~\bibnamefont {Sobierajski}}, \bibinfo {author}
  {\bibfnamefont {J.}~\bibnamefont {Krzywinski}}, \bibinfo {author}
  {\bibfnamefont {J.}~\bibnamefont {Chalupsky}}, \bibinfo {author}
  {\bibfnamefont {E.}~\bibnamefont {Abreu}}, \emph {et~al.},\ }\href
  {https://doi.org/10.1038/nphys1341} {\bibfield  {journal} {\bibinfo
  {journal} {Nat. Phys.}\ }\textbf {\bibinfo {volume} {5}},\ \bibinfo {pages}
  {693} (\bibinfo {year} {2009})}\BibitemShut {NoStop}%
\bibitem [{\citenamefont {Sch\"utte}\ \emph {et~al.}(2012)\citenamefont
  {Sch\"utte}, \citenamefont {Bauch}, \citenamefont {Fr\"uhling}, \citenamefont
  {Wieland}, \citenamefont {Gensch}, \citenamefont {Pl\"onjes}, \citenamefont
  {Gaumnitz}, \citenamefont {Azima}, \citenamefont {Bonitz},\ and\
  \citenamefont {Drescher}}]{schuette_prl_12}%
  \BibitemOpen
  \bibfield  {author} {\bibinfo {author} {\bibfnamefont {B.}~\bibnamefont
  {Sch\"utte}}, \bibinfo {author} {\bibfnamefont {S.}~\bibnamefont {Bauch}},
  \bibinfo {author} {\bibfnamefont {U.}~\bibnamefont {Fr\"uhling}}, \bibinfo
  {author} {\bibfnamefont {M.}~\bibnamefont {Wieland}}, \bibinfo {author}
  {\bibfnamefont {M.}~\bibnamefont {Gensch}}, \bibinfo {author} {\bibfnamefont
  {E.}~\bibnamefont {Pl\"onjes}}, \bibinfo {author} {\bibfnamefont
  {T.}~\bibnamefont {Gaumnitz}}, \bibinfo {author} {\bibfnamefont
  {A.}~\bibnamefont {Azima}}, \bibinfo {author} {\bibfnamefont
  {M.}~\bibnamefont {Bonitz}},\ and\ \bibinfo {author} {\bibfnamefont
  {M.}~\bibnamefont {Drescher}},\ }\href
  {https://doi.org/10.1103/PhysRevLett.108.253003} {\bibfield  {journal}
  {\bibinfo  {journal} {Phys. Rev. Lett.}\ }\textbf {\bibinfo {volume} {108}},\
  \bibinfo {pages} {253003} (\bibinfo {year} {2012})}\BibitemShut {NoStop}%
\bibitem [{\citenamefont {Almbladh}\ \emph {et~al.}(1989)\citenamefont
  {Almbladh}, \citenamefont {Morales},\ and\ \citenamefont
  {Grossmann}}]{PhysRevB.39.3489}%
  \BibitemOpen
  \bibfield  {author} {\bibinfo {author} {\bibfnamefont {C.-O.}\ \bibnamefont
  {Almbladh}}, \bibinfo {author} {\bibfnamefont {A.~L.}\ \bibnamefont
  {Morales}},\ and\ \bibinfo {author} {\bibfnamefont {G.}~\bibnamefont
  {Grossmann}},\ }\href {https://doi.org/10.1103/PhysRevB.39.3489} {\bibfield
  {journal} {\bibinfo  {journal} {Phys. Rev. B}\ }\textbf {\bibinfo {volume}
  {39}},\ \bibinfo {pages} {3489} (\bibinfo {year} {1989})}\BibitemShut
  {NoStop}%
\bibitem [{\citenamefont {Petek}\ \emph {et~al.}(1999)\citenamefont {Petek},
  \citenamefont {Nagano},\ and\ \citenamefont {Ogawa}}]{PhysRevLett.83.832}%
  \BibitemOpen
  \bibfield  {author} {\bibinfo {author} {\bibfnamefont {H.}~\bibnamefont
  {Petek}}, \bibinfo {author} {\bibfnamefont {H.}~\bibnamefont {Nagano}},\ and\
  \bibinfo {author} {\bibfnamefont {S.}~\bibnamefont {Ogawa}},\ }\href
  {https://doi.org/10.1103/PhysRevLett.83.832} {\bibfield  {journal} {\bibinfo
  {journal} {Phys. Rev. Lett.}\ }\textbf {\bibinfo {volume} {83}},\ \bibinfo
  {pages} {832} (\bibinfo {year} {1999})}\BibitemShut {NoStop}%
\bibitem [{\citenamefont {Ma}\ \emph {et~al.}(2019)\citenamefont {Ma},
  \citenamefont {Dai}, \citenamefont {Kang}, \citenamefont {Murillo},
  \citenamefont {Hou}, \citenamefont {Zhao},\ and\ \citenamefont
  {Yuan}}]{PhysRevLett.122.015001}%
  \BibitemOpen
  \bibfield  {author} {\bibinfo {author} {\bibfnamefont {Q.}~\bibnamefont
  {Ma}}, \bibinfo {author} {\bibfnamefont {J.}~\bibnamefont {Dai}}, \bibinfo
  {author} {\bibfnamefont {D.}~\bibnamefont {Kang}}, \bibinfo {author}
  {\bibfnamefont {M.~S.}\ \bibnamefont {Murillo}}, \bibinfo {author}
  {\bibfnamefont {Y.}~\bibnamefont {Hou}}, \bibinfo {author} {\bibfnamefont
  {Z.}~\bibnamefont {Zhao}},\ and\ \bibinfo {author} {\bibfnamefont
  {J.}~\bibnamefont {Yuan}},\ }\href
  {https://doi.org/10.1103/PhysRevLett.122.015001} {\bibfield  {journal}
  {\bibinfo  {journal} {Phys. Rev. Lett.}\ }\textbf {\bibinfo {volume} {122}},\
  \bibinfo {pages} {015001} (\bibinfo {year} {2019})}\BibitemShut {NoStop}%
\bibitem [{\citenamefont {Siders}\ \emph {et~al.}(1999)\citenamefont {Siders},
  \citenamefont {Cavalleri}, \citenamefont {Sokolowski-Tinten}, \citenamefont
  {T{\'o}th}, \citenamefont {Guo}, \citenamefont {Kammler}, \citenamefont
  {Hoegen}, \citenamefont {Wilson}, \citenamefont {Linde},\ and\ \citenamefont
  {Barty}}]{sokolowski_99}%
  \BibitemOpen
  \bibfield  {author} {\bibinfo {author} {\bibfnamefont {C.~W.}\ \bibnamefont
  {Siders}}, \bibinfo {author} {\bibfnamefont {A.}~\bibnamefont {Cavalleri}},
  \bibinfo {author} {\bibfnamefont {K.}~\bibnamefont {Sokolowski-Tinten}},
  \bibinfo {author} {\bibfnamefont {C.}~\bibnamefont {T{\'o}th}}, \bibinfo
  {author} {\bibfnamefont {T.}~\bibnamefont {Guo}}, \bibinfo {author}
  {\bibfnamefont {M.}~\bibnamefont {Kammler}}, \bibinfo {author} {\bibfnamefont
  {M.~H.~v.}\ \bibnamefont {Hoegen}}, \bibinfo {author} {\bibfnamefont {K.~R.}\
  \bibnamefont {Wilson}}, \bibinfo {author} {\bibfnamefont {D.~v.~d.}\
  \bibnamefont {Linde}},\ and\ \bibinfo {author} {\bibfnamefont {C.~P.~J.}\
  \bibnamefont {Barty}},\ }\href
  {https://doi.org/10.1126/science.286.5443.1340} {\bibfield  {journal}
  {\bibinfo  {journal} {Science}\ }\textbf {\bibinfo {volume} {286}},\ \bibinfo
  {pages} {1340} (\bibinfo {year} {1999})}\BibitemShut {NoStop}%
\bibitem [{\citenamefont {Chung}\ \emph {et~al.}(2017)\citenamefont {Chung},
  \citenamefont {Cho}, \citenamefont {Ciricosta}, \citenamefont {Vinko},
  \citenamefont {Wark},\ and\ \citenamefont {Lee}}]{SCFLY}%
  \BibitemOpen
  \bibfield  {author} {\bibinfo {author} {\bibfnamefont {H.-K.}\ \bibnamefont
  {Chung}}, \bibinfo {author} {\bibfnamefont {B.~I.}\ \bibnamefont {Cho}},
  \bibinfo {author} {\bibfnamefont {O.}~\bibnamefont {Ciricosta}}, \bibinfo
  {author} {\bibfnamefont {S.~M.}\ \bibnamefont {Vinko}}, \bibinfo {author}
  {\bibfnamefont {J.~S.}\ \bibnamefont {Wark}},\ and\ \bibinfo {author}
  {\bibfnamefont {R.~W.}\ \bibnamefont {Lee}},\ }\href
  {https://doi.org/10.1063/1.4975712} {\bibfield  {journal} {\bibinfo
  {journal} {AIP Conf. Proc.}\ }\textbf {\bibinfo {volume} {1811}},\ \bibinfo
  {pages} {020001} (\bibinfo {year} {2017})}\BibitemShut {NoStop}%
\bibitem [{\citenamefont {Medvedev}\ and\ \citenamefont
  {Rethfeld}(2010)}]{medvedev_10}%
  \BibitemOpen
  \bibfield  {author} {\bibinfo {author} {\bibfnamefont {N.}~\bibnamefont
  {Medvedev}}\ and\ \bibinfo {author} {\bibfnamefont {B.}~\bibnamefont
  {Rethfeld}},\ }\href {https://doi.org/10.1063/1.3511455} {\bibfield
  {journal} {\bibinfo  {journal} {J. Appl. Phys.}\ }\textbf {\bibinfo {volume}
  {108}},\ \bibinfo {pages} {103112} (\bibinfo {year} {2010})}\BibitemShut
  {NoStop}%
\bibitem [{\citenamefont {Balzer}\ \emph {et~al.}(2010)\citenamefont {Balzer},
  \citenamefont {Bauch},\ and\ \citenamefont {Bonitz}}]{balzer_pra_10_2}%
  \BibitemOpen
  \bibfield  {author} {\bibinfo {author} {\bibfnamefont {K.}~\bibnamefont
  {Balzer}}, \bibinfo {author} {\bibfnamefont {S.}~\bibnamefont {Bauch}},\ and\
  \bibinfo {author} {\bibfnamefont {M.}~\bibnamefont {Bonitz}},\ }\href
  {https://doi.org/10.1103/PhysRevA.82.033427} {\bibfield  {journal} {\bibinfo
  {journal} {Phys. Rev. A}\ }\textbf {\bibinfo {volume} {82}},\ \bibinfo
  {pages} {033427} (\bibinfo {year} {2010})}\BibitemShut {NoStop}%
\bibitem [{\citenamefont {Hermanns}\ \emph {et~al.}(2014)\citenamefont
  {Hermanns}, \citenamefont {Schl{\"u}nzen},\ and\ \citenamefont
  {Bonitz}}]{hermanns_prb14}%
  \BibitemOpen
  \bibfield  {author} {\bibinfo {author} {\bibfnamefont {S.}~\bibnamefont
  {Hermanns}}, \bibinfo {author} {\bibfnamefont {N.}~\bibnamefont
  {Schl{\"u}nzen}},\ and\ \bibinfo {author} {\bibfnamefont {M.}~\bibnamefont
  {Bonitz}},\ }\href {https://doi.org/10.1103/PhysRevB.90.125111} {\bibfield
  {journal} {\bibinfo  {journal} {Phys. Rev. B}\ }\textbf {\bibinfo {volume}
  {90}},\ \bibinfo {pages} {125111} (\bibinfo {year} {2014})}\BibitemShut
  {NoStop}%
\bibitem [{\citenamefont {Schlünzen}\ \emph {et~al.}(2020)\citenamefont
  {Schlünzen}, \citenamefont {Hermanns}, \citenamefont {Scharnke},\ and\
  \citenamefont {Bonitz}}]{schluenzen_jpcm_19}%
  \BibitemOpen
  \bibfield  {author} {\bibinfo {author} {\bibfnamefont {N.}~\bibnamefont
  {Schlünzen}}, \bibinfo {author} {\bibfnamefont {S.}~\bibnamefont
  {Hermanns}}, \bibinfo {author} {\bibfnamefont {M.}~\bibnamefont {Scharnke}},\
  and\ \bibinfo {author} {\bibfnamefont {M.}~\bibnamefont {Bonitz}},\ }\href
  {http://iopscience.iop.org/10.1088/1361-648X/ab2d32} {\bibfield  {journal}
  {\bibinfo  {journal} {J. Phys. Condens. Matter}\ }\textbf {\bibinfo {volume}
  {32}},\ \bibinfo {pages} {103001} (\bibinfo {year} {2020})}\BibitemShut
  {NoStop}%
\bibitem [{\citenamefont {Baczewski}\ \emph {et~al.}(2016)\citenamefont
  {Baczewski}, \citenamefont {Shulenburger}, \citenamefont {Desjarlais},
  \citenamefont {Hansen},\ and\ \citenamefont {Magyar}}]{Baczewski}%
  \BibitemOpen
  \bibfield  {author} {\bibinfo {author} {\bibfnamefont {A.~D.}\ \bibnamefont
  {Baczewski}}, \bibinfo {author} {\bibfnamefont {L.}~\bibnamefont
  {Shulenburger}}, \bibinfo {author} {\bibfnamefont {M.~P.}\ \bibnamefont
  {Desjarlais}}, \bibinfo {author} {\bibfnamefont {S.~B.}\ \bibnamefont
  {Hansen}},\ and\ \bibinfo {author} {\bibfnamefont {R.~J.}\ \bibnamefont
  {Magyar}},\ }\href {https://doi.org/10.1103/PhysRevLett.116.115004}
  {\bibfield  {journal} {\bibinfo  {journal} {Phys. Rev. Lett.}\ }\textbf
  {\bibinfo {volume} {116}},\ \bibinfo {pages} {115004} (\bibinfo {year}
  {2016})}\BibitemShut {NoStop}%
\bibitem [{\citenamefont {Mo}\ \emph {et~al.}(2018)\citenamefont {Mo},
  \citenamefont {Fu}, \citenamefont {Kang}, \citenamefont {Zhang},\ and\
  \citenamefont {He}}]{PhysRevLett.120.205002}%
  \BibitemOpen
  \bibfield  {author} {\bibinfo {author} {\bibfnamefont {C.}~\bibnamefont
  {Mo}}, \bibinfo {author} {\bibfnamefont {Z.}~\bibnamefont {Fu}}, \bibinfo
  {author} {\bibfnamefont {W.}~\bibnamefont {Kang}}, \bibinfo {author}
  {\bibfnamefont {P.}~\bibnamefont {Zhang}},\ and\ \bibinfo {author}
  {\bibfnamefont {X.~T.}\ \bibnamefont {He}},\ }\href
  {https://doi.org/10.1103/PhysRevLett.120.205002} {\bibfield  {journal}
  {\bibinfo  {journal} {Phys. Rev. Lett.}\ }\textbf {\bibinfo {volume} {120}},\
  \bibinfo {pages} {205002} (\bibinfo {year} {2018})}\BibitemShut {NoStop}%
\bibitem [{\citenamefont {Runge}\ and\ \citenamefont {Gross}(1984)}]{TDDFT}%
  \BibitemOpen
  \bibfield  {author} {\bibinfo {author} {\bibfnamefont {E.}~\bibnamefont
  {Runge}}\ and\ \bibinfo {author} {\bibfnamefont {E.~K.~U.}\ \bibnamefont
  {Gross}},\ }\href {https://doi.org/10.1103/PhysRevLett.52.997} {\bibfield
  {journal} {\bibinfo  {journal} {Phys. Rev. Lett.}\ }\textbf {\bibinfo
  {volume} {52}},\ \bibinfo {pages} {997} (\bibinfo {year} {1984})}\BibitemShut
  {NoStop}%
\bibitem [{\citenamefont {Silaeva}\ \emph {et~al.}(2018)\citenamefont
  {Silaeva}, \citenamefont {Bevillon}, \citenamefont {Stoian},\ and\
  \citenamefont {Colombier}}]{silaeva2018}%
  \BibitemOpen
  \bibfield  {author} {\bibinfo {author} {\bibfnamefont {E.~P.}\ \bibnamefont
  {Silaeva}}, \bibinfo {author} {\bibfnamefont {E.}~\bibnamefont {Bevillon}},
  \bibinfo {author} {\bibfnamefont {R.}~\bibnamefont {Stoian}},\ and\ \bibinfo
  {author} {\bibfnamefont {J.~P.}\ \bibnamefont {Colombier}},\ }\href
  {https://doi.org/10.1103/PhysRevB.98.094306} {\bibfield  {journal} {\bibinfo
  {journal} {Phys. Rev. B}\ }\textbf {\bibinfo {volume} {98}},\ \bibinfo
  {pages} {094306} (\bibinfo {year} {2018})}\BibitemShut {NoStop}%
\bibitem [{\citenamefont {Zhang}\ \emph
  {et~al.}(2016{\natexlab{b}})\citenamefont {Zhang}, \citenamefont {Zhao},
  \citenamefont {Kang}, \citenamefont {Zhang},\ and\ \citenamefont
  {He}}]{PhysRevB.93.115114}%
  \BibitemOpen
  \bibfield  {author} {\bibinfo {author} {\bibfnamefont {S.}~\bibnamefont
  {Zhang}}, \bibinfo {author} {\bibfnamefont {S.}~\bibnamefont {Zhao}},
  \bibinfo {author} {\bibfnamefont {W.}~\bibnamefont {Kang}}, \bibinfo {author}
  {\bibfnamefont {P.}~\bibnamefont {Zhang}},\ and\ \bibinfo {author}
  {\bibfnamefont {X.-T.}\ \bibnamefont {He}},\ }\href
  {https://doi.org/10.1103/PhysRevB.93.115114} {\bibfield  {journal} {\bibinfo
  {journal} {Phys. Rev. B}\ }\textbf {\bibinfo {volume} {93}},\ \bibinfo
  {pages} {115114} (\bibinfo {year} {2016}{\natexlab{b}})}\BibitemShut
  {NoStop}%
\bibitem [{\citenamefont {Dai}\ \emph {et~al.}(2010)\citenamefont {Dai},
  \citenamefont {Hou},\ and\ \citenamefont {Yuan}}]{PhysRevLett.104.245001}%
  \BibitemOpen
  \bibfield  {author} {\bibinfo {author} {\bibfnamefont {J.}~\bibnamefont
  {Dai}}, \bibinfo {author} {\bibfnamefont {Y.}~\bibnamefont {Hou}},\ and\
  \bibinfo {author} {\bibfnamefont {J.}~\bibnamefont {Yuan}},\ }\href
  {https://doi.org/10.1103/PhysRevLett.104.245001} {\bibfield  {journal}
  {\bibinfo  {journal} {Phys. Rev. Lett.}\ }\textbf {\bibinfo {volume} {104}},\
  \bibinfo {pages} {245001} (\bibinfo {year} {2010})}\BibitemShut {NoStop}%
\bibitem [{\citenamefont {Dai}\ \emph {et~al.}(2017)\citenamefont {Dai},
  \citenamefont {Gao}, \citenamefont {Sun},\ and\ \citenamefont
  {Kang}}]{Dai_2017}%
  \BibitemOpen
  \bibfield  {author} {\bibinfo {author} {\bibfnamefont {J.}~\bibnamefont
  {Dai}}, \bibinfo {author} {\bibfnamefont {C.}~\bibnamefont {Gao}}, \bibinfo
  {author} {\bibfnamefont {H.}~\bibnamefont {Sun}},\ and\ \bibinfo {author}
  {\bibfnamefont {D.}~\bibnamefont {Kang}},\ }\href
  {https://doi.org/10.1088/1361-6455/aa84f8} {\bibfield  {journal} {\bibinfo
  {journal} {J. Phys. B}\ }\textbf {\bibinfo {volume} {50}},\ \bibinfo {pages}
  {184004} (\bibinfo {year} {2017})}\BibitemShut {NoStop}%
\bibitem [{\citenamefont {Vinko}\ \emph {et~al.}(2014)\citenamefont {Vinko},
  \citenamefont {Ciricosta},\ and\ \citenamefont {Wark}}]{RN100}%
  \BibitemOpen
  \bibfield  {author} {\bibinfo {author} {\bibfnamefont {S.~M.}\ \bibnamefont
  {Vinko}}, \bibinfo {author} {\bibfnamefont {O.}~\bibnamefont {Ciricosta}},\
  and\ \bibinfo {author} {\bibfnamefont {J.~S.}\ \bibnamefont {Wark}},\ }\href
  {https://doi.org/10.1038/ncomms4533} {\bibfield  {journal} {\bibinfo
  {journal} {Nat. Commun.}\ }\textbf {\bibinfo {volume} {5}},\ \bibinfo {pages}
  {3533} (\bibinfo {year} {2014})}\BibitemShut {NoStop}%
\bibitem [{\citenamefont {Taillefumier}\ \emph {et~al.}(2002)\citenamefont
  {Taillefumier}, \citenamefont {Cabaret}, \citenamefont {Flank},\ and\
  \citenamefont {Mauri}}]{taillefumier2002}%
  \BibitemOpen
  \bibfield  {author} {\bibinfo {author} {\bibfnamefont {M.}~\bibnamefont
  {Taillefumier}}, \bibinfo {author} {\bibfnamefont {D.}~\bibnamefont
  {Cabaret}}, \bibinfo {author} {\bibfnamefont {A.-M.}\ \bibnamefont {Flank}},\
  and\ \bibinfo {author} {\bibfnamefont {F.}~\bibnamefont {Mauri}},\ }\href
  {https://doi.org/10.1103/PhysRevB.66.195107} {\bibfield  {journal} {\bibinfo
  {journal} {Phys. Rev. B}\ }\textbf {\bibinfo {volume} {66}},\ \bibinfo
  {pages} {195107} (\bibinfo {year} {2002})}\BibitemShut {NoStop}%
\bibitem [{\citenamefont {Kang}\ \emph {et~al.}(2013)\citenamefont {Kang},
  \citenamefont {Dai}, \citenamefont {Sun}, \citenamefont {Hou},\ and\
  \citenamefont {Yuan}}]{RN98}%
  \BibitemOpen
  \bibfield  {author} {\bibinfo {author} {\bibfnamefont {D.}~\bibnamefont
  {Kang}}, \bibinfo {author} {\bibfnamefont {J.}~\bibnamefont {Dai}}, \bibinfo
  {author} {\bibfnamefont {H.}~\bibnamefont {Sun}}, \bibinfo {author}
  {\bibfnamefont {Y.}~\bibnamefont {Hou}},\ and\ \bibinfo {author}
  {\bibfnamefont {J.}~\bibnamefont {Yuan}},\ }\href
  {https://doi.org/10.1038/srep03272} {\bibfield  {journal} {\bibinfo
  {journal} {Sci. Rep.}\ }\textbf {\bibinfo {volume} {3}},\ \bibinfo {pages}
  {3272} (\bibinfo {year} {2013})}\BibitemShut {NoStop}%
\bibitem [{\citenamefont {Celliers}\ \emph {et~al.}(2018)\citenamefont
  {Celliers}, \citenamefont {Millot}, \citenamefont {Brygoo}, \citenamefont
  {McWilliams}, \citenamefont {Fratanduono}, \citenamefont {Rygg},
  \citenamefont {Goncharov}, \citenamefont {Loubeyre}, \citenamefont {Eggert},
  \citenamefont {Peterson} \emph {et~al.}}]{Celliers677}%
  \BibitemOpen
  \bibfield  {author} {\bibinfo {author} {\bibfnamefont {P.~M.}\ \bibnamefont
  {Celliers}}, \bibinfo {author} {\bibfnamefont {M.}~\bibnamefont {Millot}},
  \bibinfo {author} {\bibfnamefont {S.}~\bibnamefont {Brygoo}}, \bibinfo
  {author} {\bibfnamefont {R.~S.}\ \bibnamefont {McWilliams}}, \bibinfo
  {author} {\bibfnamefont {D.~E.}\ \bibnamefont {Fratanduono}}, \bibinfo
  {author} {\bibfnamefont {J.~R.}\ \bibnamefont {Rygg}}, \bibinfo {author}
  {\bibfnamefont {A.~F.}\ \bibnamefont {Goncharov}}, \bibinfo {author}
  {\bibfnamefont {P.}~\bibnamefont {Loubeyre}}, \bibinfo {author}
  {\bibfnamefont {J.~H.}\ \bibnamefont {Eggert}}, \bibinfo {author}
  {\bibfnamefont {J.~L.}\ \bibnamefont {Peterson}}, \emph {et~al.},\ }\href
  {https://doi.org/10.1126/science.aat0970} {\bibfield  {journal} {\bibinfo
  {journal} {Science}\ }\textbf {\bibinfo {volume} {361}},\ \bibinfo {pages}
  {677} (\bibinfo {year} {2018})}\BibitemShut {NoStop}%
\bibitem [{\citenamefont {Schattke}\ and\ \citenamefont {van
  Hove}(2003)}]{schattke-book}%
  \BibitemOpen
  \bibfield  {author} {\bibinfo {author} {\bibfnamefont {W.}~\bibnamefont
  {Schattke}}\ and\ \bibinfo {author} {\bibfnamefont {M.}~\bibnamefont {van
  Hove}},\ }\href@noop {} {\emph {\bibinfo {title} {{Solid-State Photoemission
  and Related Methods}}}}\ (\bibinfo  {publisher} {Wiley VCH},\ \bibinfo {year}
  {2003})\BibitemShut {NoStop}%
\bibitem [{\citenamefont {Winter}\ and\ \citenamefont
  {Aumayr}(1999)}]{aumayr_99}%
  \BibitemOpen
  \bibfield  {author} {\bibinfo {author} {\bibfnamefont {H.}~\bibnamefont
  {Winter}}\ and\ \bibinfo {author} {\bibfnamefont {F.}~\bibnamefont
  {Aumayr}},\ }\href {https://doi.org/10.1088/0953-4075/32/7/005} {\bibfield
  {journal} {\bibinfo  {journal} {J. Phys. B}\ }\textbf {\bibinfo {volume}
  {32}},\ \bibinfo {pages} {R39} (\bibinfo {year} {1999})}\BibitemShut
  {NoStop}%
\bibitem [{\citenamefont {Anisimov}\ \emph {et~al.}(1974)\citenamefont
  {Anisimov}, \citenamefont {Kapeliovich}, \citenamefont {Perelman} \emph
  {et~al.}}]{anisimov1974electron}%
  \BibitemOpen
  \bibfield  {author} {\bibinfo {author} {\bibfnamefont {S.~I.}\ \bibnamefont
  {Anisimov}}, \bibinfo {author} {\bibfnamefont {B.~L.}\ \bibnamefont
  {Kapeliovich}}, \bibinfo {author} {\bibfnamefont {T.~L.}\ \bibnamefont
  {Perelman}}, \emph {et~al.},\ }\href@noop {} {\bibfield  {journal} {\bibinfo
  {journal} {Sov. Phys. JETP}\ }\textbf {\bibinfo {volume} {39}},\ \bibinfo
  {pages} {375} (\bibinfo {year} {1974})}\BibitemShut {NoStop}%
\bibitem [{\citenamefont {Zeng}\ and\ \citenamefont {Dai}(2020)}]{RN99}%
  \BibitemOpen
  \bibfield  {author} {\bibinfo {author} {\bibfnamefont {Q.}~\bibnamefont
  {Zeng}}\ and\ \bibinfo {author} {\bibfnamefont {J.}~\bibnamefont {Dai}},\
  }\href {https://doi.org/10.1007/s11433-019-1466-2} {\bibfield  {journal}
  {\bibinfo  {journal} {Sci. China Phys. Mech.}\ }\textbf {\bibinfo {volume}
  {63}},\ \bibinfo {pages} {263011} (\bibinfo {year} {2020})}\BibitemShut
  {NoStop}%
\bibitem [{\citenamefont {Lin}\ \emph {et~al.}(2008)\citenamefont {Lin},
  \citenamefont {Zhigilei},\ and\ \citenamefont {Celli}}]{PhysRevB.77.075133}%
  \BibitemOpen
  \bibfield  {author} {\bibinfo {author} {\bibfnamefont {Z.}~\bibnamefont
  {Lin}}, \bibinfo {author} {\bibfnamefont {L.~V.}\ \bibnamefont {Zhigilei}},\
  and\ \bibinfo {author} {\bibfnamefont {V.}~\bibnamefont {Celli}},\ }\href
  {https://doi.org/10.1103/PhysRevB.77.075133} {\bibfield  {journal} {\bibinfo
  {journal} {Phys. Rev. B}\ }\textbf {\bibinfo {volume} {77}},\ \bibinfo
  {pages} {075133} (\bibinfo {year} {2008})}\BibitemShut {NoStop}%
\bibitem [{\citenamefont {Giannozzi}\ \emph {et~al.}(2009)\citenamefont
  {Giannozzi}, \citenamefont {Baroni}, \citenamefont {Bonini}, \citenamefont
  {Calandra}, \citenamefont {Car}, \citenamefont {Cavazzoni}, \citenamefont
  {Ceresoli}, \citenamefont {Chiarotti}, \citenamefont {Cococcioni},
  \citenamefont {Dabo} \emph {et~al.}}]{giannozzi2009}%
  \BibitemOpen
  \bibfield  {author} {\bibinfo {author} {\bibfnamefont {P.}~\bibnamefont
  {Giannozzi}}, \bibinfo {author} {\bibfnamefont {S.}~\bibnamefont {Baroni}},
  \bibinfo {author} {\bibfnamefont {N.}~\bibnamefont {Bonini}}, \bibinfo
  {author} {\bibfnamefont {M.}~\bibnamefont {Calandra}}, \bibinfo {author}
  {\bibfnamefont {R.}~\bibnamefont {Car}}, \bibinfo {author} {\bibfnamefont
  {C.}~\bibnamefont {Cavazzoni}}, \bibinfo {author} {\bibfnamefont
  {D.}~\bibnamefont {Ceresoli}}, \bibinfo {author} {\bibfnamefont {G.~L.}\
  \bibnamefont {Chiarotti}}, \bibinfo {author} {\bibfnamefont {M.}~\bibnamefont
  {Cococcioni}}, \bibinfo {author} {\bibfnamefont {I.}~\bibnamefont {Dabo}},
  \emph {et~al.},\ }\href {https://doi.org/10.1088/0953-8984/21/39/395502}
  {\bibfield  {journal} {\bibinfo  {journal} {J. Phys. Condens. Matter}\
  }\textbf {\bibinfo {volume} {21}},\ \bibinfo {pages} {395502} (\bibinfo
  {year} {2009})}\BibitemShut {NoStop}%
\bibitem [{\citenamefont {Giannozzi}\ \emph {et~al.}(2017)\citenamefont
  {Giannozzi}, \citenamefont {Andreussi}, \citenamefont {Brumme}, \citenamefont
  {Bunau}, \citenamefont {Nardelli}, \citenamefont {Calandra}, \citenamefont
  {Car}, \citenamefont {Cavazzoni}, \citenamefont {Ceresoli}, \citenamefont
  {Cococcioni} \emph {et~al.}}]{Giannozzi_2017}%
  \BibitemOpen
  \bibfield  {author} {\bibinfo {author} {\bibfnamefont {P.}~\bibnamefont
  {Giannozzi}}, \bibinfo {author} {\bibfnamefont {O.}~\bibnamefont
  {Andreussi}}, \bibinfo {author} {\bibfnamefont {T.}~\bibnamefont {Brumme}},
  \bibinfo {author} {\bibfnamefont {O.}~\bibnamefont {Bunau}}, \bibinfo
  {author} {\bibfnamefont {M.~B.}\ \bibnamefont {Nardelli}}, \bibinfo {author}
  {\bibfnamefont {M.}~\bibnamefont {Calandra}}, \bibinfo {author}
  {\bibfnamefont {R.}~\bibnamefont {Car}}, \bibinfo {author} {\bibfnamefont
  {C.}~\bibnamefont {Cavazzoni}}, \bibinfo {author} {\bibfnamefont
  {D.}~\bibnamefont {Ceresoli}}, \bibinfo {author} {\bibfnamefont
  {M.}~\bibnamefont {Cococcioni}}, \emph {et~al.},\ }\href
  {https://doi.org/10.1088/1361-648x/aa8f79} {\bibfield  {journal} {\bibinfo
  {journal} {J. Phys. Condens. Matter}\ }\textbf {\bibinfo {volume} {29}},\
  \bibinfo {pages} {465901} (\bibinfo {year} {2017})}\BibitemShut {NoStop}%
\bibitem [{\citenamefont {Kohn}\ and\ \citenamefont {Sham}(1965)}]{KohnSham}%
  \BibitemOpen
  \bibfield  {author} {\bibinfo {author} {\bibfnamefont {W.}~\bibnamefont
  {Kohn}}\ and\ \bibinfo {author} {\bibfnamefont {L.~J.}\ \bibnamefont
  {Sham}},\ }\href {https://doi.org/10.1103/PhysRev.140.A1133} {\bibfield
  {journal} {\bibinfo  {journal} {Phys. Rev.}\ }\textbf {\bibinfo {volume}
  {140}},\ \bibinfo {pages} {A1133} (\bibinfo {year} {1965})}\BibitemShut
  {NoStop}%
\bibitem [{\citenamefont {Mermin}(1965)}]{Mermin}%
  \BibitemOpen
  \bibfield  {author} {\bibinfo {author} {\bibfnamefont {N.~D.}\ \bibnamefont
  {Mermin}},\ }\href {https://doi.org/10.1103/PhysRev.137.A1441} {\bibfield
  {journal} {\bibinfo  {journal} {Phys. Rev.}\ }\textbf {\bibinfo {volume}
  {137}},\ \bibinfo {pages} {A1441} (\bibinfo {year} {1965})}\BibitemShut
  {NoStop}%
\bibitem [{\citenamefont {Bl\"ochl}(1994)}]{PAW}%
  \BibitemOpen
  \bibfield  {author} {\bibinfo {author} {\bibfnamefont {P.~E.}\ \bibnamefont
  {Bl\"ochl}},\ }\href {https://doi.org/10.1103/PhysRevB.50.17953} {\bibfield
  {journal} {\bibinfo  {journal} {Phys. Rev. B}\ }\textbf {\bibinfo {volume}
  {50}},\ \bibinfo {pages} {17953} (\bibinfo {year} {1994})}\BibitemShut
  {NoStop}%
\bibitem [{\citenamefont {Perdew}\ and\ \citenamefont
  {Zunger}(1981)}]{PerdewZunger}%
  \BibitemOpen
  \bibfield  {author} {\bibinfo {author} {\bibfnamefont {J.~P.}\ \bibnamefont
  {Perdew}}\ and\ \bibinfo {author} {\bibfnamefont {A.}~\bibnamefont
  {Zunger}},\ }\href {https://doi.org/10.1103/PhysRevB.23.5048} {\bibfield
  {journal} {\bibinfo  {journal} {Phys. Rev. B}\ }\textbf {\bibinfo {volume}
  {23}},\ \bibinfo {pages} {5048} (\bibinfo {year} {1981})}\BibitemShut
  {NoStop}%
\bibitem [{\citenamefont {Groth}\ \emph {et~al.}(2017)\citenamefont {Groth},
  \citenamefont {Dornheim}, \citenamefont {Sjostrom}, \citenamefont {Malone},
  \citenamefont {Foulkes},\ and\ \citenamefont {Bonitz}}]{groth_prl17}%
  \BibitemOpen
  \bibfield  {author} {\bibinfo {author} {\bibfnamefont {S.}~\bibnamefont
  {Groth}}, \bibinfo {author} {\bibfnamefont {T.}~\bibnamefont {Dornheim}},
  \bibinfo {author} {\bibfnamefont {T.}~\bibnamefont {Sjostrom}}, \bibinfo
  {author} {\bibfnamefont {F.~D.}\ \bibnamefont {Malone}}, \bibinfo {author}
  {\bibfnamefont {W.~M.~C.}\ \bibnamefont {Foulkes}},\ and\ \bibinfo {author}
  {\bibfnamefont {M.}~\bibnamefont {Bonitz}},\ }\href
  {https://doi.org/10.1103/PhysRevLett.119.135001} {\bibfield  {journal}
  {\bibinfo  {journal} {Phys. Rev. Lett.}\ }\textbf {\bibinfo {volume} {119}},\
  \bibinfo {pages} {135001} (\bibinfo {year} {2017})}\BibitemShut {NoStop}%
\bibitem [{\citenamefont {Monkhorst}\ and\ \citenamefont
  {Pack}(1976)}]{MPgrid}%
  \BibitemOpen
  \bibfield  {author} {\bibinfo {author} {\bibfnamefont {H.~J.}\ \bibnamefont
  {Monkhorst}}\ and\ \bibinfo {author} {\bibfnamefont {J.~D.}\ \bibnamefont
  {Pack}},\ }\href {https://doi.org/10.1103/PhysRevB.13.5188} {\bibfield
  {journal} {\bibinfo  {journal} {Phys. Rev. B}\ }\textbf {\bibinfo {volume}
  {13}},\ \bibinfo {pages} {5188} (\bibinfo {year} {1976})}\BibitemShut
  {NoStop}%
\bibitem [{\citenamefont {Kresse}\ and\ \citenamefont {Hafner}(1994)}]{AIMD}%
  \BibitemOpen
  \bibfield  {author} {\bibinfo {author} {\bibfnamefont {G.}~\bibnamefont
  {Kresse}}\ and\ \bibinfo {author} {\bibfnamefont {J.}~\bibnamefont
  {Hafner}},\ }\href {https://doi.org/10.1103/PhysRevB.49.14251} {\bibfield
  {journal} {\bibinfo  {journal} {Phys. Rev. B}\ }\textbf {\bibinfo {volume}
  {49}},\ \bibinfo {pages} {14251} (\bibinfo {year} {1994})}\BibitemShut
  {NoStop}%
\bibitem [{\citenamefont {Calderín}\ \emph {et~al.}(2017)\citenamefont
  {Calderín}, \citenamefont {Karasiev},\ and\ \citenamefont
  {Trickey}}]{calderin2017}%
  \BibitemOpen
  \bibfield  {author} {\bibinfo {author} {\bibfnamefont {L.}~\bibnamefont
  {Calderín}}, \bibinfo {author} {\bibfnamefont {V.}~\bibnamefont
  {Karasiev}},\ and\ \bibinfo {author} {\bibfnamefont {S.}~\bibnamefont
  {Trickey}},\ }\href
  {https://doi.org/https://doi.org/10.1016/j.cpc.2017.08.008} {\bibfield
  {journal} {\bibinfo  {journal} {Comput. Phys. Commun.}\ }\textbf {\bibinfo
  {volume} {221}},\ \bibinfo {pages} {118 } (\bibinfo {year}
  {2017})}\BibitemShut {NoStop}%
\bibitem [{\citenamefont {Henke}\ \emph {et~al.}(1993)\citenamefont {Henke},
  \citenamefont {Gullikson},\ and\ \citenamefont {Davis}}]{henke1993x}%
  \BibitemOpen
  \bibfield  {author} {\bibinfo {author} {\bibfnamefont {B.}~\bibnamefont
  {Henke}}, \bibinfo {author} {\bibfnamefont {E.}~\bibnamefont {Gullikson}},\
  and\ \bibinfo {author} {\bibfnamefont {J.}~\bibnamefont {Davis}},\ }\href
  {https://doi.org/https://doi.org/10.1006/adnd.1993.1013} {\bibfield
  {journal} {\bibinfo  {journal} {At. Data Nucl. Data Tables}\ }\textbf
  {\bibinfo {volume} {54}},\ \bibinfo {pages} {181 } (\bibinfo {year}
  {1993})}\BibitemShut {NoStop}%
\bibitem [{\citenamefont {Yuan}(2002)}]{YUAN2002275}%
  \BibitemOpen
  \bibfield  {author} {\bibinfo {author} {\bibfnamefont {J.~M.}\ \bibnamefont
  {Yuan}},\ }\href
  {https://doi.org/https://doi.org/10.1016/S0368-2048(01)00365-6} {\bibfield
  {journal} {\bibinfo  {journal} {J. Electron Spectros. Relat. Phenomena}\
  }\textbf {\bibinfo {volume} {122}},\ \bibinfo {pages} {275 } (\bibinfo {year}
  {2002})}\BibitemShut {NoStop}%
\bibitem [{\citenamefont {Inglis}\ and\ \citenamefont
  {Teller}(1939)}]{inglisteller}%
  \BibitemOpen
  \bibfield  {author} {\bibinfo {author} {\bibfnamefont {D.~R.}\ \bibnamefont
  {Inglis}}\ and\ \bibinfo {author} {\bibfnamefont {E.}~\bibnamefont
  {Teller}},\ }\href@noop {} {\bibfield  {journal} {\bibinfo  {journal}
  {Astrophys. J.}\ }\textbf {\bibinfo {volume} {90}},\ \bibinfo {pages} {439}
  (\bibinfo {year} {1939})}\BibitemShut {NoStop}%
\bibitem [{\citenamefont {Ciricosta}\ \emph {et~al.}(2012)\citenamefont
  {Ciricosta}, \citenamefont {Vinko}, \citenamefont {Chung}, \citenamefont
  {Cho}, \citenamefont {Brown}, \citenamefont {Burian}, \citenamefont
  {Chalupsk\'y}, \citenamefont {Engelhorn}, \citenamefont {Falcone},
  \citenamefont {Graves} \emph {et~al.}}]{PhysRevLett.109.065002}%
  \BibitemOpen
  \bibfield  {author} {\bibinfo {author} {\bibfnamefont {O.}~\bibnamefont
  {Ciricosta}}, \bibinfo {author} {\bibfnamefont {S.~M.}\ \bibnamefont
  {Vinko}}, \bibinfo {author} {\bibfnamefont {H.-K.}\ \bibnamefont {Chung}},
  \bibinfo {author} {\bibfnamefont {B.-I.}\ \bibnamefont {Cho}}, \bibinfo
  {author} {\bibfnamefont {C.~R.~D.}\ \bibnamefont {Brown}}, \bibinfo {author}
  {\bibfnamefont {T.}~\bibnamefont {Burian}}, \bibinfo {author} {\bibfnamefont
  {J.}~\bibnamefont {Chalupsk\'y}}, \bibinfo {author} {\bibfnamefont
  {K.}~\bibnamefont {Engelhorn}}, \bibinfo {author} {\bibfnamefont {R.~W.}\
  \bibnamefont {Falcone}}, \bibinfo {author} {\bibfnamefont {C.}~\bibnamefont
  {Graves}}, \emph {et~al.},\ }\href
  {https://doi.org/10.1103/PhysRevLett.109.065002} {\bibfield  {journal}
  {\bibinfo  {journal} {Phys. Rev. Lett.}\ }\textbf {\bibinfo {volume} {109}},\
  \bibinfo {pages} {065002} (\bibinfo {year} {2012})}\BibitemShut {NoStop}%
\bibitem [{\citenamefont {Hoarty}\ \emph {et~al.}(2013)\citenamefont {Hoarty},
  \citenamefont {Allan}, \citenamefont {James}, \citenamefont {Brown},
  \citenamefont {Hobbs}, \citenamefont {Hill}, \citenamefont {Harris},
  \citenamefont {Morton}, \citenamefont {Brookes}, \citenamefont {Shepherd}
  \emph {et~al.}}]{PhysRevLett.110.265003}%
  \BibitemOpen
  \bibfield  {author} {\bibinfo {author} {\bibfnamefont {D.~J.}\ \bibnamefont
  {Hoarty}}, \bibinfo {author} {\bibfnamefont {P.}~\bibnamefont {Allan}},
  \bibinfo {author} {\bibfnamefont {S.~F.}\ \bibnamefont {James}}, \bibinfo
  {author} {\bibfnamefont {C.~R.~D.}\ \bibnamefont {Brown}}, \bibinfo {author}
  {\bibfnamefont {L.~M.~R.}\ \bibnamefont {Hobbs}}, \bibinfo {author}
  {\bibfnamefont {M.~P.}\ \bibnamefont {Hill}}, \bibinfo {author}
  {\bibfnamefont {J.~W.~O.}\ \bibnamefont {Harris}}, \bibinfo {author}
  {\bibfnamefont {J.}~\bibnamefont {Morton}}, \bibinfo {author} {\bibfnamefont
  {M.~G.}\ \bibnamefont {Brookes}}, \bibinfo {author} {\bibfnamefont
  {R.}~\bibnamefont {Shepherd}}, \emph {et~al.},\ }\href
  {https://doi.org/10.1103/PhysRevLett.110.265003} {\bibfield  {journal}
  {\bibinfo  {journal} {Phys. Rev. Lett.}\ }\textbf {\bibinfo {volume} {110}},\
  \bibinfo {pages} {265003} (\bibinfo {year} {2013})}\BibitemShut {NoStop}%
\bibitem [{\citenamefont {Mathewson}\ and\ \citenamefont
  {Myers}(1972)}]{doi:10.1080/14786437208229308}%
  \BibitemOpen
  \bibfield  {author} {\bibinfo {author} {\bibfnamefont {A.~G.}\ \bibnamefont
  {Mathewson}}\ and\ \bibinfo {author} {\bibfnamefont {H.~P.}\ \bibnamefont
  {Myers}},\ }\href {https://doi.org/10.1080/14786437208229308} {\bibfield
  {journal} {\bibinfo  {journal} {Phil. Mag.}\ }\textbf {\bibinfo {volume}
  {25}},\ \bibinfo {pages} {853} (\bibinfo {year} {1972})}\BibitemShut
  {NoStop}%
\bibitem [{\citenamefont {Mathewson}\ and\ \citenamefont
  {Myers}(1971)}]{Mathewson_1971}%
  \BibitemOpen
  \bibfield  {author} {\bibinfo {author} {\bibfnamefont {A.~G.}\ \bibnamefont
  {Mathewson}}\ and\ \bibinfo {author} {\bibfnamefont {H.~P.}\ \bibnamefont
  {Myers}},\ }\href {https://doi.org/10.1088/0031-8949/4/6/009} {\bibfield
  {journal} {\bibinfo  {journal} {Phys. Scr.}\ }\textbf {\bibinfo {volume}
  {4}},\ \bibinfo {pages} {291} (\bibinfo {year} {1971})}\BibitemShut {NoStop}%
\bibitem [{\citenamefont {Johnson}\ and\ \citenamefont
  {Christy}(1975)}]{PhysRevB.11.1315}%
  \BibitemOpen
  \bibfield  {author} {\bibinfo {author} {\bibfnamefont {P.~B.}\ \bibnamefont
  {Johnson}}\ and\ \bibinfo {author} {\bibfnamefont {R.~W.}\ \bibnamefont
  {Christy}},\ }\href {https://doi.org/10.1103/PhysRevB.11.1315} {\bibfield
  {journal} {\bibinfo  {journal} {Phys. Rev. B}\ }\textbf {\bibinfo {volume}
  {11}},\ \bibinfo {pages} {1315} (\bibinfo {year} {1975})}\BibitemShut
  {NoStop}%
\bibitem [{\citenamefont {Johnson}\ and\ \citenamefont
  {Christy}(1972)}]{PhysRevB.6.4370}%
  \BibitemOpen
  \bibfield  {author} {\bibinfo {author} {\bibfnamefont {P.~B.}\ \bibnamefont
  {Johnson}}\ and\ \bibinfo {author} {\bibfnamefont {R.~W.}\ \bibnamefont
  {Christy}},\ }\href {https://doi.org/10.1103/PhysRevB.6.4370} {\bibfield
  {journal} {\bibinfo  {journal} {Phys. Rev. B}\ }\textbf {\bibinfo {volume}
  {6}},\ \bibinfo {pages} {4370} (\bibinfo {year} {1972})}\BibitemShut
  {NoStop}%
\bibitem [{\citenamefont {Wang}\ \emph {et~al.}(2013)\citenamefont {Wang},
  \citenamefont {Long}, \citenamefont {Tian}, \citenamefont {He},\ and\
  \citenamefont {Zhang}}]{congwang2013}%
  \BibitemOpen
  \bibfield  {author} {\bibinfo {author} {\bibfnamefont {C.}~\bibnamefont
  {Wang}}, \bibinfo {author} {\bibfnamefont {Y.}~\bibnamefont {Long}}, \bibinfo
  {author} {\bibfnamefont {M.-F.}\ \bibnamefont {Tian}}, \bibinfo {author}
  {\bibfnamefont {X.-T.}\ \bibnamefont {He}},\ and\ \bibinfo {author}
  {\bibfnamefont {P.}~\bibnamefont {Zhang}},\ }\href
  {https://doi.org/10.1103/PhysRevE.87.043105} {\bibfield  {journal} {\bibinfo
  {journal} {Phys. Rev. E}\ }\textbf {\bibinfo {volume} {87}},\ \bibinfo
  {pages} {043105} (\bibinfo {year} {2013})}\BibitemShut {NoStop}%
\bibitem [{\citenamefont {Holst}\ \emph {et~al.}(2014)\citenamefont {Holst},
  \citenamefont {Recoules}, \citenamefont {Mazevet}, \citenamefont {Torrent},
  \citenamefont {Ng}, \citenamefont {Chen}, \citenamefont {Kirkwood},
  \citenamefont {Sametoglu}, \citenamefont {Reid},\ and\ \citenamefont
  {Tsui}}]{PhysRevB.90.035121}%
  \BibitemOpen
  \bibfield  {author} {\bibinfo {author} {\bibfnamefont {B.}~\bibnamefont
  {Holst}}, \bibinfo {author} {\bibfnamefont {V.}~\bibnamefont {Recoules}},
  \bibinfo {author} {\bibfnamefont {S.}~\bibnamefont {Mazevet}}, \bibinfo
  {author} {\bibfnamefont {M.}~\bibnamefont {Torrent}}, \bibinfo {author}
  {\bibfnamefont {A.}~\bibnamefont {Ng}}, \bibinfo {author} {\bibfnamefont
  {Z.}~\bibnamefont {Chen}}, \bibinfo {author} {\bibfnamefont {S.~E.}\
  \bibnamefont {Kirkwood}}, \bibinfo {author} {\bibfnamefont {V.}~\bibnamefont
  {Sametoglu}}, \bibinfo {author} {\bibfnamefont {M.}~\bibnamefont {Reid}},\
  and\ \bibinfo {author} {\bibfnamefont {Y.~Y.}\ \bibnamefont {Tsui}},\ }\href
  {https://doi.org/10.1103/PhysRevB.90.035121} {\bibfield  {journal} {\bibinfo
  {journal} {Phys. Rev. B}\ }\textbf {\bibinfo {volume} {90}},\ \bibinfo
  {pages} {035121} (\bibinfo {year} {2014})}\BibitemShut {NoStop}%
\bibitem [{\citenamefont {Ceperley}\ and\ \citenamefont
  {Alder}(1980)}]{PhysRevLett.45.566}%
  \BibitemOpen
  \bibfield  {author} {\bibinfo {author} {\bibfnamefont {D.~M.}\ \bibnamefont
  {Ceperley}}\ and\ \bibinfo {author} {\bibfnamefont {B.~J.}\ \bibnamefont
  {Alder}},\ }\href {https://doi.org/10.1103/PhysRevLett.45.566} {\bibfield
  {journal} {\bibinfo  {journal} {Phys. Rev. Lett.}\ }\textbf {\bibinfo
  {volume} {45}},\ \bibinfo {pages} {566} (\bibinfo {year} {1980})}\BibitemShut
  {NoStop}%
\bibitem [{\citenamefont {Dornheim}\ \emph {et~al.}(2018)\citenamefont
  {Dornheim}, \citenamefont {Groth},\ and\ \citenamefont
  {Bonitz}}]{dornheim_physrep_18}%
  \BibitemOpen
  \bibfield  {author} {\bibinfo {author} {\bibfnamefont {T.}~\bibnamefont
  {Dornheim}}, \bibinfo {author} {\bibfnamefont {S.}~\bibnamefont {Groth}},\
  and\ \bibinfo {author} {\bibfnamefont {M.}~\bibnamefont {Bonitz}},\ }\href
  {https://doi.org/10.1016/j.physrep.2018.04.001} {\bibfield  {journal}
  {\bibinfo  {journal} {Phys. Rep.}\ }\textbf {\bibinfo {volume} {744}},\
  \bibinfo {pages} {1 } (\bibinfo {year} {2018})}\BibitemShut {NoStop}%
\bibitem [{\citenamefont {Schoof}\ \emph {et~al.}(2015)\citenamefont {Schoof},
  \citenamefont {Groth}, \citenamefont {Vorberger},\ and\ \citenamefont
  {Bonitz}}]{schoof_prl15}%
  \BibitemOpen
  \bibfield  {author} {\bibinfo {author} {\bibfnamefont {T.}~\bibnamefont
  {Schoof}}, \bibinfo {author} {\bibfnamefont {S.}~\bibnamefont {Groth}},
  \bibinfo {author} {\bibfnamefont {J.}~\bibnamefont {Vorberger}},\ and\
  \bibinfo {author} {\bibfnamefont {M.}~\bibnamefont {Bonitz}},\ }\href
  {https://doi.org/10.1103/PhysRevLett.115.130402} {\bibfield  {journal}
  {\bibinfo  {journal} {Phys. Rev. Lett.}\ }\textbf {\bibinfo {volume} {115}},\
  \bibinfo {pages} {130402} (\bibinfo {year} {2015})}\BibitemShut {NoStop}%
\bibitem [{\citenamefont {Dornheim}\ \emph {et~al.}(2016)\citenamefont
  {Dornheim}, \citenamefont {Groth}, \citenamefont {Sjostrom}, \citenamefont
  {Malone}, \citenamefont {Foulkes},\ and\ \citenamefont
  {Bonitz}}]{dornheim_prl16}%
  \BibitemOpen
  \bibfield  {author} {\bibinfo {author} {\bibfnamefont {T.}~\bibnamefont
  {Dornheim}}, \bibinfo {author} {\bibfnamefont {S.}~\bibnamefont {Groth}},
  \bibinfo {author} {\bibfnamefont {T.}~\bibnamefont {Sjostrom}}, \bibinfo
  {author} {\bibfnamefont {F.~D.}\ \bibnamefont {Malone}}, \bibinfo {author}
  {\bibfnamefont {W.~M.~C.}\ \bibnamefont {Foulkes}},\ and\ \bibinfo {author}
  {\bibfnamefont {M.}~\bibnamefont {Bonitz}},\ }\href
  {https://doi.org/10.1103/PhysRevLett.117.156403} {\bibfield  {journal}
  {\bibinfo  {journal} {Phys. Rev. Lett.}\ }\textbf {\bibinfo {volume} {117}},\
  \bibinfo {pages} {156403} (\bibinfo {year} {2016})}\BibitemShut {NoStop}%
\bibitem [{\citenamefont {Karasiev}\ \emph {et~al.}(2016)\citenamefont
  {Karasiev}, \citenamefont {Calder\'{\i}n},\ and\ \citenamefont
  {Trickey}}]{PhysRevE.93.063207}%
  \BibitemOpen
  \bibfield  {author} {\bibinfo {author} {\bibfnamefont {V.~V.}\ \bibnamefont
  {Karasiev}}, \bibinfo {author} {\bibfnamefont {L.}~\bibnamefont
  {Calder\'{\i}n}},\ and\ \bibinfo {author} {\bibfnamefont {S.~B.}\
  \bibnamefont {Trickey}},\ }\href {https://doi.org/10.1103/PhysRevE.93.063207}
  {\bibfield  {journal} {\bibinfo  {journal} {Phys. Rev. E}\ }\textbf {\bibinfo
  {volume} {93}},\ \bibinfo {pages} {063207} (\bibinfo {year}
  {2016})}\BibitemShut {NoStop}%
\bibitem [{\citenamefont {Kwong}\ \emph {et~al.}(1998)\citenamefont {Kwong},
  \citenamefont {Bonitz}, \citenamefont {Binder},\ and\ \citenamefont
  {K\"ohler}}]{kwong-etal.98pss}%
  \BibitemOpen
  \bibfield  {author} {\bibinfo {author} {\bibfnamefont {N.~H.}\ \bibnamefont
  {Kwong}}, \bibinfo {author} {\bibfnamefont {M.}~\bibnamefont {Bonitz}},
  \bibinfo {author} {\bibfnamefont {R.}~\bibnamefont {Binder}},\ and\ \bibinfo
  {author} {\bibfnamefont {H.~S.}\ \bibnamefont {K\"ohler}},\ }\href
  {https://doi.org/10.1002/(SICI)1521-3951(199803)206:1<197::AID-PSSB197>3.0.CO;2-9}
  {\bibfield  {journal} {\bibinfo  {journal} {Phys. Stat. Sol. B}\ }\textbf
  {\bibinfo {volume} {206}},\ \bibinfo {pages} {197} (\bibinfo {year}
  {1998})}\BibitemShut {NoStop}%
\end{thebibliography}%

\end{document}